\documentclass[twocolumn]{aastex631}

\newcommand{\hii}{{\sc H\,ii}}
\newcommand{\oi}{[{\sc O\,i}]}
\newcommand{\oii}{[{\sc O\,ii}]}
\newcommand{\oiii}{[{\sc O\,iii}]}
\newcommand{\nii}{[{\sc N\,ii}]}
\newcommand{\sii}{[{\sc S\,ii}]}
\newcommand{\luci}{\texttt{LUCI}}
\usepackage{pythonhighlight}
\usepackage{xcolor} 

\received{\today}
\revised{\today}
\accepted{\today}
\shorttitle{NGC 1275 Optical-Emission Nebula}
\shortauthors{Rhea et al.}
\begin{document}

\title{Mapping the Filamentary Nebula of NGC 1275 with Multiwavelength SITELLE Observations}

\author[0000-0003-2001-1076]{Carter Lee Rhea}
\affiliation{D\'{e}partement de Physique, Universit\'{e} de Montr\'{e}al, Succ. Centre-Ville, Montr\'{e}al, Qu\'{e}bec, H3C 3J7, Canada}
\affiliation{Centre de recherche en astrophysique du Québec (CRAQ)}
\affiliation{Dragonfly Focused Research Organization, 150 Washington Avenue, Santa Fe, 87501, NM, USA}

\author[0000-0001-7271-7340]{Julie Hlavacek-Larrondo}
\affiliation{D\'{e}partement de Physique, Universit\'{e} de Montr\'{e}al, Succ. Centre-Ville, Montr\'{e}al, Qu\'{e}bec, H3C 3J7, Canada}

\author[0000-0002-2478-5119]{Marie-Lou Gendron-Marsolais}
\affiliation{Instituto de Astrofísica de Andalucía, IAA-CSIC, Apartado 3004, 18080 Granada, España}
\affiliation{Département de physique, de génie physique et d’optique, Université Laval, Québec (QC), G1V 0A6, Canada}

\author[0000-0002-2478-5119]{Benjamin Vigneron}
\affiliation{D\'{e}partement de Physique, Universit\'{e} de Montr\'{e}al, Succ. Centre-Ville, Montr\'{e}al, Qu\'{e}bec, H3C 3J7, Canada}

\author[0000-0002-2808-0853]{Megan Donahue}
\affiliation{Michigan State University, Department of Physics and Astronomy, East Lansing, MI 48824, USA}

\author[0000-0001-5223-1888]{Auriane Thilloy} 
\affiliation{D\'{e}partement de Physique, Universit\'{e} de Montr\'{e}al, Succ. Centre-Ville, Montr\'{e}al, Qu\'{e}bec, H3C 3J7, Canada}

\author[0000-0002-5136-6673]{Laurie Rousseau-Nepton}
\affiliation{Canada-France-Hawaii Telescope, 65-1238 Mamalahoa Hwy, Kamuela, Hawaii 96743, USA}
\affiliation{David A. Dunlap Department of Astronomy \& Astrophysics, University of Toronto, 50 St. George Street, Toronto, ON M5S 3H4, Canada}
\affiliation{Dunlap Institute for Astronomy \& Astrophysics, University of Toronto, 50 St. George Street, Toronto, ON M5S 3H4, Canada}

\author[0000-0003-4440-259X]{Mar Mezcua}
\affiliation{Institute of Space Sciences (ICE, CSIC), Campus UAB, Carrer de Magrans, 08193 Barcelona, Spain}
\affiliation{Institut d'Estudis Espacials de Catalunya (IEEC), Edifici RDIT, Campus UPC, 08860 Castelldefels (Barcelona), Spain}

\author[0000-0003-0392-0120]{Norbert Werner}
\affiliation{Department of Theoretical Physics and Astrophysics, Faculty of Science, Masaryk University, Kotl\'a\v rsk\'a 2, Brno, 611 37, Czech Republic}

\author[0000-0003-2405-7258]{Jorge Barrera-Ballesteros}
\affiliation{ Universidad Nacional Aut\'onoma de M\'exico, Instituto de Astronom\'ia, AP 70-264, CDMX 04510, M\'exico}

\author[0000-0002-3173-1098]{Hyunseop Choi}
\affiliation{Département de Physique, Université de Montréal, Succ. Centre-Ville, Montréal, Québec, H3C 3J7, Canada}
\affiliation{Mila - Quebec Artificial Intelligence Institute, Montreal, Quebec, Canada}

\author[0000-0002-3398-6916]{Alastair Edge}
\affiliation{Centre for Extragalactic Astronomy, Department of Physics, Durham University, South Road, Durham DH1 3LE, UK}

\author{Andrew Fabian}
\affiliation{Institute of Astronomy, Madingley road, Cambridge CB3 0HA, UK}

\author[ 0000-0002-3514-0383 ]{G. Mark Voit}
\affiliation{Michigan State University, Department of Physics and Astronomy, East Lansing, MI 48824, USA}

\begin{abstract}
The filamentary nebula encompassing the central galaxy of the Perseus Cluster, NGC 1275, is a complex structure extending dozens of kiloparsecs from NGC 1275. Decades of previous works have focused on establishing the primary formation and ionization mechanisms in different filaments. These studies have pointed to a lack of star formation in the majority of the filaments, the importance of magnetic fields and turbulence in several regions, and the role of interactions between the intercluster medium (ICM) and the cool gas in the filaments, as well as the role of interaction between the central radio source, 3C84, and the filaments. In this paper, we present multi-filter observations of the entire filamentary system that cover the optical bandpass, using the SITELLE instrument at the Canada-France-Hawai'i Telescope. 
Here, we use the data analysis software, \href{https://crhea93.github.io/LUCI/index.html}{\texttt{LUCI}}, to produce flux maps of the prominent emission lines present in the filters: \oii{}$\lambda$3726/3729, \oiii{}$\lambda$5007, H$\beta$, \nii{}$\lambda$6548, \nii{}$\lambda$6583, and H$\alpha$. We use these maps to produce BPT and WHAN diagrams to study the ionization mechanisms at play in each distinct region of the filamentary nebula. First, we confirm the absence of \oiii{}$\lambda$5007 in the extended filaments, although we detect this line in the central core, revealing a compact region where photoionization by the AGN might affect local conditions.
Our findings corroborate previous claims that the ionization in the extended filaments could be caused by the cooling ICM via collisional excitation and/or mixing. Moreover, they support the conclusion that magnetic fields play an important role in the formation and continued existence of the filaments.

\end{abstract}

\keywords{Galaxies : NGC 1275  --- Supermassive Black Hole (SMBH)  ---  intracluster medium (ICM)  --- filamentary nebula }

\section{Introduction} \label{sec:intro}

Recent observational campaigns targeting nearby galaxies have revealed the presence of a diffuse halo of gas surrounding their host galaxies known as the Circum Galactic Medium (CGM; see \citealt{tumlinson_circumgalactic_2017}, \citealt{suresh_impact_2015}, and \citealt{barai_galactic_2013} for details). The CGM contains the majority of baryons in the galactic system and is thus of vital importance in our understanding of galaxy evolution (e.g., \citealt{bregman_extended_2018}; \citealt{macquart_census_2020}). 
However, due to the diffuse nature of the gas, the CGM is only detectable in systems with rare extremely deep observations or those that fortuitously have an illuminating background source (e.g., \citealt{steidel_structure_2010}; \citealt{zhu_calcium_2013}).

Despite the fact that the CGM is difficult to study in the majority of galaxies with contemporary observatories, extremely massive galaxies are expected to have very large reservoirs of this diffuse gas. These massive galaxies live in the center of galaxy clusters ($\ge$ 100 galaxies) and have been shown to host a denser, though still diffuse, halo of gas known as the Intra Cluster Medium (ICM) which acts as an excellent analog to the CGM of individual galaxies. Indeed, observations have demonstrated that the relative abundances of baryonic matter in the form of stars and gas in addition to dark matter in the ICM are consistent with those of the CGM (e.g., \citealt{marrone_locuss_2012}). Therefore, studying the ICM, readily detectable with current instruments, is a good starting point in order to glean properties of the CGM.

%ICM is observed as multiphase gas surrounding the central galaxy of galaxy clusters (e.g., \citealt{minkowki_optical_1959}).
Multiwavelength observations of the ICM around the nearest and brightest galaxy clusters have revealed that the ICM is highly multiphase \citealt{minkowki_optical_1959}), spanning several orders of magnitude in temperature and density. Therefore, it can be detected in the X-rays through the highly ionized hot inter cluster medium ($10^{7-8}$ K), H$\alpha$ ($10^4$ K) and/or in the infrared ($\sim30$ K). Moreover, certain nebulae have been shown to host thread-like filaments stretching over several kiloparsecs (e.g., \citealt{fabian_magnetic_2008}; \citealt{mcdonald_optical_2011}).

The origin of the multiphase gas in the ICM is still unknown. Although optical emission of this type is associated with star formation, there is little evidence of this in the ICM (e.g., \citealt{fabian_chandra_2000}; \citealt{canning_star_2010}; \citealt{mcdonald_optical_2011}).
This multiphase gas could form \textit{in situ} through a self-regulated feedback loop in which the central black hole heats the surrounding medium, while producing a wealth of thermal instabilities that cause some gas to cool, condense and rain back down on the central black hole, re-igniting the feedback loop (e.g., see the \textit{precipitation} in \citealt{voit_precipitation-regulated_2015}, \textit{cold feedback} in \citealt{pizzolato_nature_2005} and \textit{chaotic cold accretion} in \citealt{gaspari_chaotic_2013}). The multiphase gas could also originate from cold gas that has been uplifted by the radio jets of the central black hole (e.g., \citealt{mcnamara_1010_2014}) or as a result of both processes, as the low-entropy gas that has been uplifted by the radio jet becomes thermally unstable when it is lifted to higher altitudes (see \textit{stimulated feedback} model in  \citealt{mcnamara_mechanism_2016}). 

In this work, we target the most extensively studied cluster of galaxies with multiphase gas in the literature, the Perseus cluster, located at a redshift of z$=$0.017284 (\citealt{hitomi_collaboration_quiescent_2016}; \citealt{fabian_chandra_2000}; \citealt{fabian_heating_2006}; \citealt{conselice_nature_2001}; \citealt{canning_star_2010}; \citealt{hatch_origin_2006}; \citealt{salome_cold_2008}). Due to its proximity and extensive multiwavelength coverage, we studied 
the optically emitting cool gas in this target with exquisite spatial resolution (116 pc/pixel).

Perseus also exhibits powerful radio jets that are co-spatial with cavities as seen at X-ray wavelengths in the ICM (e.g., \citealt{mushotzky_observation_1981}; \citealt{fabian_properties_2002}; \citealt{schmidt_chandra_2002}; \citealt{churazov_xmm-newton_2003}; \citealt{sanders_mapping_2004}). Several studies indicate that the cooling of the ICM, which would result in a cooling flow, is being halted by the injection of energy. This energy is introduced by jets emanating from the active galactic nucleus (AGN) of the central galaxy from the large scales to the small scales through a process of turbulent cascade of energy. Consequently, making the cluster a prototype for radio mode feedback (e.g., \citealt{fabian_heating_2006}). Moreover, these types of clusters with AGN feedback have been found to host filamentary nebula of cool gas (e.g., \citealt{mcdonald_optical_2011}).

Indeed, the central dominant galaxy, NGC 1275 is surrounded by a spectacular nebula of multiphase gas, first reported in \cite{minkowki_optical_1959}. This work, and subsequent imaging, revealed the size of the nebula, which covers an area of 80 kpc $\times$ 50 kpc corresponding to 218 arcseconds $\times$ 136 arcseconds. 
Follow up observations reveal the presence of an interloping galaxy, refered as the High-Velocity System (HVS), seen in projection and that is believed to be infalling on NGC 1275. In velocity space, the HVS is offset by an impressive 3000 km/s with respect to the central galaxy NGC 1275 (e.g., \citealt{minkowki_optical_1959}; \citealt{lynds_improved_1970}; \citealt{rubin_new_1977}; \citealt{conselice_nature_2001}; \citealt{yu_high-velocity_2015}).

\begin{figure*}
    \includegraphics[width=0.95\textwidth]{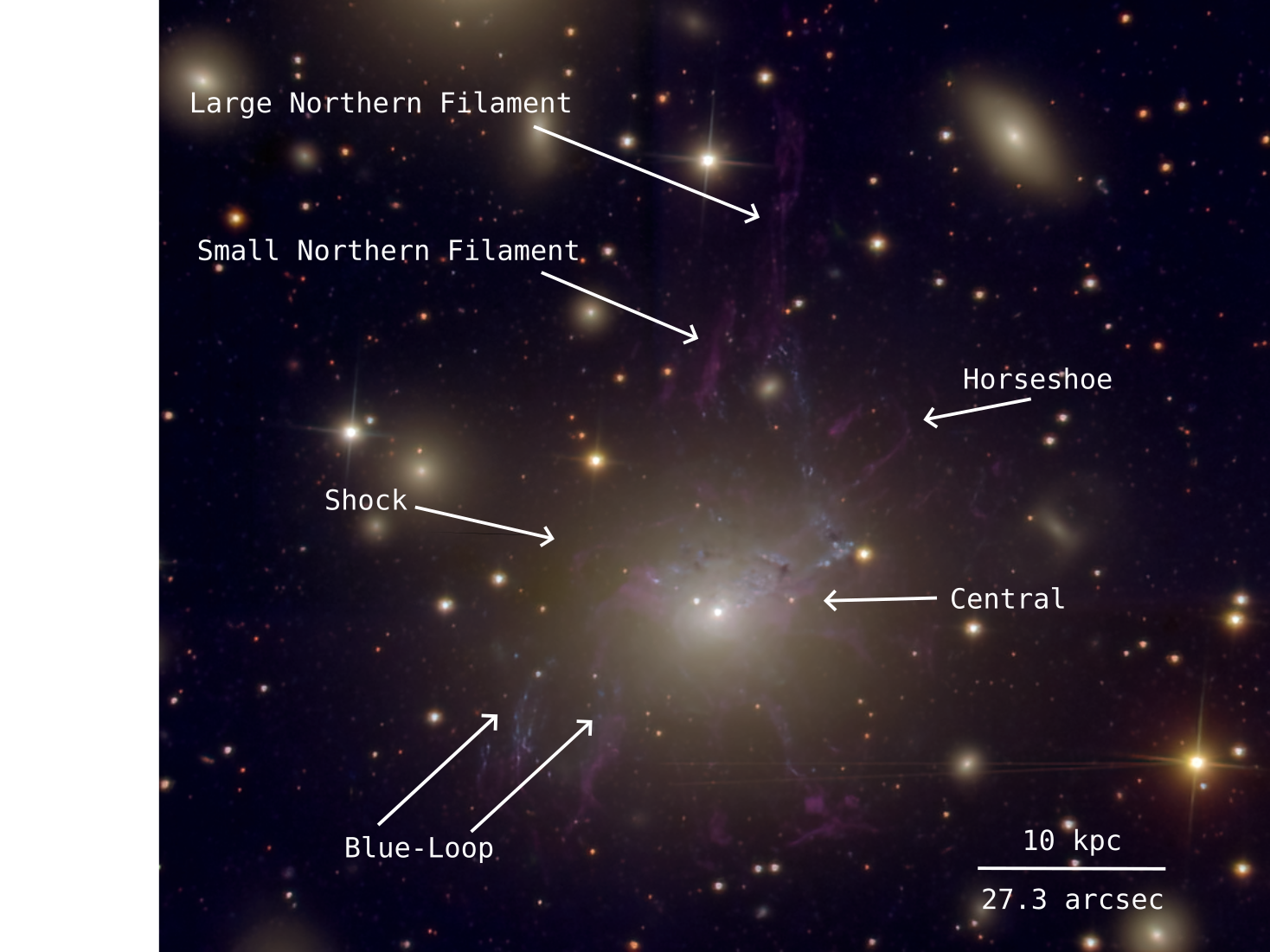}
    \caption{Composite image of the SN1 (365-385 nm; blue), SN2 (480-520 nm; green), and SN3 (651-685; red) filters emission in NGC 1275 obtained with SITELLE on the CFHT. We constrain the image only to cover approximately a quarter of the SITELLE field of view in order to highlight the optical emission nebula.}
    \label{fig:combinedImage}
\end{figure*}

The multiphase nebula in Perseus is associated with warm ionized gas ($10^4$ K ), seen through optical emission lines such as H$\alpha$ and \nii{}$\lambda$6583. The \textit{Hubble Space Telescope} (HST) observations additionally revealed that certain filamentary structures (such as the large and small Northern filament; See Figure \ref{fig:combinedImage}) are extremely thin, on the order of 70 pc in diameter, while extending 6 kpc in length (\citealt{fabian_magnetic_2008}). Importantly, these results suggest the presence of strong magnetic fields that stabilize the ionized gas filaments. Follow up UV observations taken with the HST confirmed an overall lack of young stars indicating that the origin of the nebula is not dominated by star-formation, except for small knots in the northern filament and in the region called the blue-loop (\citealt{canning_star_2010}). In light of the HVS, this filamentary nebula of gas at the same redshift of NGC 1275 is often referred to as the low velocity system (LVS).

In \cite{burbidge_ionized_1962}, the authors observed that the \nii{} to H$\alpha$ ratio was near unity in NGC 1275 and argued that photoionization by O and B stars was insufficient to explain this higher-than-usual ratio. 
Subsequent works found similar results (e.g., \citealt{hatch_origin_2006}; \citealt{gendron-marsolais_revealing_2018}; \citealt{vigneron_high-spectral-resolution_2024})
Rather, a ratio of unity between these lines is more prominent in the nuclei  of galaxies (e.g., \citealt{dickey_agn_2019}; \citealt{oh_observational_2019}; \citealt{davies_starburst-agn_2014}). Interestingly, consistent values around unity of the \nii{}/H$\alpha$ ratio are found well away from the AGN of NGC 1275.

The role of the AGN in ionizing the extended, optically emitting cool filaments surrounding NGC 1275 has been a subject of debate. Decades of observations have demonstrated the relatively limited influence of its activity as an ionizing source, rather suggesting a possible role in affecting only the inner few kpcs of the filamentary structure. The first dedicated spectroscopic studies of NGC 1275's filaments originally proposed that shocks and the central AGN radiation could potentially ionize the optical filaments (\citealt{kent_ionization_1979}). However, subsequent spectroscopic observations led to the exploration of multiple other potential ionization mechanisms. Additionally, these analyses further concluded that NGC 1275’s AGN would be unable to ionize the optical filamentary nebula and that a combination of other ionization sources could explain the filaments emission profiles (see \citealt{heckman_dynamical_1989}, \citealt{sabra_emission-line_2000} and \citealt{conselice_nature_2001}). Nevertheless, more recent spectroscopic studies of the cool filaments surrounding NGC 1275 tentatively suggested that the central regions close to the BCG could be partly affected by the AGN's activity, or a similarly hard excitation source (see \citealt{hatch_origin_2006} and \citealt{gendron-marsolais_revealing_2018}).

The kinematics of the multiphase nebula were first measured by \cite{conselice_nature_2001} using the Densepak fiber array at the WIYN observatory\footnote{The WIYN observatory hosts a 3.5 meter telescope and is located at the Kitt Peak National Observatory in Southern Arizona}.
These measurements revealed a complex velocity structure of the warm H$\alpha$-emitting gas in the center of NGC 1275. Carefully selected slit spectroscopy along the filamentary structure in NGC 1275 by the Gemini Multi-Object Spectrometer further demonstrated the complexity of the kinematics of the system (\citealt{hatch_origin_2006}). The first complete view of the kinematics of NGC 1275 were published by \cite{gendron-marsolais_revealing_2018}; these observations were taken by the SITELLE instrument located at the Canada-France-Hawaii Telescope (CFHT). They revealed no organized rotation in the nebula. Follow-up high spectral-resolution observations with SITELLE presented in \cite{vigneron_high-spectral-resolution_2024} revealed the presence of a disk-like central structure characterised by a high velocity dispersion ($\ge$ 80 km/s). This structure also appears spatially correlated and comoving with a similar central disk of cold molecular gas as observed by \citealt{lim_radially_2008} with the Submillimeter Array. In contrast, the rest of the filamentary nebula displays a relatively low and homogeneous velocity dispersion (20-40 km/s), potentially suggesting that multiple mechanisms might be at work to produce the multiphase gas within this galaxy cluster.

Observations taken with the \textit{Chandra X-ray Observatory} also highlight the presence of the same filamentary structures in the soft-band covering 0.5 - 2.0 keV (e.g., \citealt{fabian_chandra_2000}; \citealt{fabian_heating_2006}; \citealt{fabian_relationship_2003}).
This co-spatiality between the warm H$\alpha$ emitting gas and the hot X-ray emitting ICM has been confirmed through several papers exploring the multiphase gas (e.g., \citealt{fabian_relationship_2003}; \citealt{lim_radially_2008}; \citealt{li_direct_2020}; \citealt{mcdonald_optical_2011}). It has also been extended to CO(2-1) emission tracing the 20-500 K molecular gas (\citealt{salome_cold_2008}; \citealt{nagai_alma_2019}) and far infrared red atomic gas transitions such as \nii{}122$\mu$m tracing cold gas between 40 and 120 K (\citealt{mittal_herschel_2012}). 

%We constrain the image only to cover approximately a quarter of the SITELLE field of view in order to highlight the optical-emission nebula. Here, we highlight the main structures of interest.

In this work, we present new additional SITELLE flux maps, targeting the \oiii{}$\lambda$5007, H$\beta$, and \oii{}$\lambda$3726+\oii{}$\lambda$3729 emission line. These observations cover for the first time the entire filamentary nebula in NGC 1275 at these wavelengths. We use these maps to investigate the ionization mechanism of the nebula using standard and modified Baldwin-Phillips-Terlevich (BPT) diagrams (e.g., \citealt{baldwin_classification_1981}). As discussed throughout the paper, it is difficult to fully constrain the ionization mechanism for the filaments using the BPT diagram.
In $\S$2, we present the observations and the data analysis methods. In $\S$3, we present the flux maps for the SN1, SN2, and SN3 filters observations in addition to the BPT diagrams. In $\S$4, we explore individual regions of the nebula in detail and compare them with multiwavelength studies. Finally, we will summarize our results in $\S$5 and discuss future work.

We assume a redshift of $z=0.017284$ for LVS of NGC 1275, corresponding to 21.1 kpc per arcmin (\citealt{hitomi_collaboration_hitomi_2018}; \citealt{wright_cosmology_2006}). Assuming the Hubble constant of $H_0$=69.6, the energy density of matter, $\Omega_M$=0.286, the energy density of the vacuum, $\Omega_{vac}$=0.714, and a flat curvature, this redshift is equivalent to a luminosity distance of 75.5 Mpc.

\begin{deluxetable*}{cccccc}[t!]
    \tablecaption{\label{table:SitelleObs}}
    %\tablewidth{700pt}
    \tabletypesize{\scriptsize}
    \tablehead{
        \colhead{Filter} & \colhead{Wavelength (\AA)} & \colhead{Emission Lines} &            \colhead{Resolution: $R=\frac{\Delta \lambda}{\lambda}$} & \colhead{Observation ID} & \colhead{P.I.} 
    } 
    \decimalcolnumbers
    \startdata
        SN1    & 3650-3850  &     \oii{}3276, \oii{}3279           & 1800       & 17BC22 & Gendron-Marsolais         \\ \hline
        SN2    & 4800-5200  & H$\beta$, \oiii{}4959, \oiii{}5007           & 1800       & 17BC22 & Gendron-Marsolais         \\ \hline
        SN3    & 6510-6850  & H$\alpha$, \nii{}6548, \nii{}6583, \sii{}6716, \sii{}6731, \oi{}6364          & 1800       & 16BQ12 & Morrison \\ \hline
    \enddata
    \tablecomments{Description of SITELLE filters used in this study including the observation IDs and Principal Investigator (P.I.) indicated on the proposal. We note that the SN3 data was part of the science verification stage.}
\end{deluxetable*}

\section{Observations and data reduction} \label{sec:methods}
%ADD Chandra section 
\subsection{SITELLE at the CFHT}
% Description of SITELLE data

NGC 1275 was observed in 2017 with the SN1 (3650-3850 \AA) and SN2 (4800-5200 \AA) filters with a spectral resolution of R$\approx$1800 (P.I. Gendron-Marsolais). The SN1 filter targets the \oii{}$\lambda$3726,3729 emission line doublet, while the SN2 filter captures \oiii{}$\lambda$4959, \oiii{}$\lambda$5007, and H$\beta\lambda$4861. As for the SN3 (6510-6850 \AA) filter, NGC 1275 was initially observed as part of the science verification of SITELLE with a spectral resolution R$\approx$1800 (P.I. Morrison). Although NGC 1275 has since been reobserved at a higher resolution (R$\approx$7000; P.I. Rhea $\&$ Hlavacek-Larrondo 20DD99; \citealt{vigneron_high-spectral-resolution_2024}), we use here the initial data since it has the same spectral resolution as the other two cubes. The SN3 filter targets several prominent emission lines: H$\alpha\lambda$6563, \nii{}$\lambda$6548, \nii{}$\lambda$6583, \sii{}$\lambda$6716, \sii{}$\lambda$6731, and \oi{}$\lambda$6364. 
A summary of the filters, emission lines targeted, resolution, observation ID, and P.I.s can be found in table \ref{table:SitelleObs}.
We show a composite image in figure \ref{fig:combinedImage} highlighting the regions of interest.
Each pixel of SITELLE corresponds to a projected size of approximately 113 pc $\times$ 113 pc (0.32 arcsec $\times$ 0.32 arcsec) on the sky at the redshift of NGC 1275. 

The data for all three filters were reduced and calibrated using the ORBS data pipeline (\citealt{martin_calibrations_2017}). The data analysis presented in this paper was completed using \luci{} (\citealt{rhea_luci_2021}; \citealt{rhea_crhea93luci_2021}). \texttt{LUCI} is a general-purpose line-fitting algorithm conceived specifically for SITELLE. \texttt{LUCI} uses machine learning algorithms trained to predict the velocity and velocity dispersion of emission lines present in a SITELLE spectrum to initialize an optimization algorithm that fits a model to the observed spectrum. Following \cite{martin_optimal_2016}, we fit a sinc function convolved with a Gaussian function to each emission line. 
The sinc-gauss function is written as follows:
\begin{equation}
    \text{SG(x)} = A e^{-b^2} \frac{\text{erf}(a-ib)+\text{erf}(a+ib)}{2 \text{erf}(a)}
\end{equation}
where 
\begin{equation}
    a = \frac{\Delta \sigma}{\sqrt{2}\Delta w}
\end{equation}
and 
\begin{equation}
    b = \frac{x-x_0}{\sqrt{2}\Delta \sigma}
\end{equation}.
$x$ is the position, $x_0$ is the rest position of the line, $A$ is the maximum value, $\Delta w$ is the width of the sinc function, and $\Delta \sigma$ is the width of the Gaussian function.

Since we provide the redshift of the filamentary nebula to \texttt{LUCI}, the fitting process excludes the HVS.

The continuum is fitted assuming a zeroth order polynomial. \texttt{LUCI} utilizes the \textit{scipy.optimization.minimize} implementation of the Sequential Least SQuares Programming (SLSQP) algorithm. The machine learning algorithm in \texttt{LUCI} was trained on synthetic data covering a Doppler velocity of -500 km/s to 500 km/s and a broadening ranging from 10 km/s to 300 km/s. These bounds were chosen based on the kinematic maps previously obtained by \cite{gendron-marsolais_revealing_2018}.
For more details on the machine learning algorithm and synthetic spectra, we direct the reader to \cite{rhea_machine-learning_2020} and \href{https://sitelle-signals.github.io/Pamplemousse/index.html}{Pamplemousse}. Instead of obtaining point estimates from a convolutional neural network, we use the mixture density network described in \cite{rhea_updates_2021} to obtain error estimates on the line position and broadening. These are then passed into the fitting algorithm as priors to constrain the fits. This process is described in detail in \cite{rhea_updates_2021}.

Each filter was fit separately with a spatial binning of 3$\times$3 ($\approx 350$ kpc $\times 350$ kpc). The binning was chosen to maximize the detection of the emission lines, especially those present in SN1 and SN2 which are typically fainter than those present in SN3. 
%Within each filter, the emission lines were all fit simultaneously along with the continuum.
For the SN3 and SN2 filters, the velocity and velocity dispersion of the emission lines in the same filter were linked; in doing this, we explicitly assume that the same gas is responsible for these emission lines. %We do not tie the velocity and velocity dispersion of the H$\beta$ line and the \oiii{} doublet in SN2 since NGC 1275 is known to have almost no detectable \oiii{} emission (\citealt{hatch_origin_2006}). Therefore, this does not force the assumption that the same gas emitting in H$\beta$ emits in \oiii{}. At the same time, it does not prohibit the velocity and velocity dispersions to be equal between the lines. 
In SN1, each emission line's velocity and velocity dispersion are tied together for the \oii{} doublet, but due to the low spectral resolution, we are unable to disentangle the two components of the \oii{} doublet (i.e. $\lambda$3726,3729) and, thus, the flux reported is the summed flux of the two lines.

Before fitting the lines, we subtract the sky contribution from each filter.
The sky spectra are taken from a 9.6-arcsecond region centered at (\textit{3:19:42.4979, +41:32:01.810}); this region contains neither point sources nor source emission. We show the sky spectra and fits in figure \ref{fig:illustritiveFits}.
The \texttt{LUCI} fit commands used can be found in appendix \ref{app:fitCommands}.
% Background Method
%In this paper, we utilize the background modeling method described in \carter{Rhea et al. 2023b}. The background is modeled by segmenting the deep image into source and background regions using a standard segmentation algorithm. The background spaxels are then decomposed using principal component analysis (PCA) such that a linear combination of principal components plus a mean component describes the spectrum in each pixel. During the PCA process, each pixel is assigned a series of values corresponding to the coefficients of each principal component. We then train a neural network to interpolate these coefficient vectors from the background spaxels onto the previously masked source spaxels. This allows us to create a model for the background that differs from pixel to pixel. The segmentation map, coefficient maps, and component plots are shown in appendix \carter{APPENDIX}. Once calculated, this background spectrum can be subtracted from the source region's spectrum prior to fitting. We note that for SN1 and SN2, the principal components were strongly dominated by noise; therefore, we only used the mean spectrum calculated as part of the PCA.
After fitting each datacube, we then apply an alignment and masking procedure on the fitted maps. Since the astrometry of each reduced SITELLE datacube is imperfect, we align the SN1 and SN2 observations to SN3 using \texttt{astroalign.apply\_transform} (\citealt{beroiz_astroalign_2020}). We supply three manually selected reference stars present in all three filters. Finally, we verify the alignment with \textit{Chandra} observations. To align the observations, we use the position of the AGN and two other bright sources present in both the SITELLE deep images and the \textit{Chandra} observation.
%\texttt{reproject} module command \texttt{reproject\_exact}. %The command applies a flux-conserving spherical polygon intersection algorithm to the data to align the datacubes\footnote{Details of the algorithm can be found at \url{https://reproject.readthedocs.io/en/stable/celestial.html}}. 

After the alignment, we apply a mask on each emission-line map to highlight the nebular emission in final maps. The mask consist of a flux cut of 1$\times10^{-17}$ergs s$^{-1}$ cm$^{-2}$ \AA$^{-1}$ on all maps. Additionally, we apply a signal-to-noise ratio (SNR) cut of 3 (e.g., \citealt{tremblay_galaxy-scale_2018}; \citealt{rousseau-nepton_signals_2019}). We calculate the SNR after sky subtraction\footnote{We use the following command \texttt{LUCI.create\_SNR\_map(method=1)}. The signal is calculated as the peak flux in the spectrum while the noise is sampled from a region of the spectrum not containing any emission lines.}. The SNR was calculated for each line except for the \nii{}-doublet. For the \nii{}-doublet, we assumed the same SNR as H$\alpha$ since the lines are crowded in the same region of the spectra. 
Finally, we apply an additional mask to the \oiii{}$\lambda$5007 map since we needed to exclude the spaxels that overlap with the position of the HVS. Indeed, the \oiii{}$\lambda$5007 emission from the nebula is unfortunately heavily contaminated by the \oiii{}$\lambda$4959 emission from the HVS (\citealt{hatch_origin_2006}) due to the velocity shift between the systems (The \oiii{}$\lambda$4959 emission from the HVS is located at approximately 5097 \AA while the \oiii{}$\lambda$5007 emission from the nebula is at 5094 \AA). This contamination is due to the spatial and spectral overlap of these lines. The position of the HVS was determined by fitting the H$\alpha$ in the HVS in the SN2 filter.
The resulting flux maps for the emission lines in each filter are shown in figure \ref{fig:fluxMaps}.

%In order to establish the most accurate redshift to supply to \luci{} for fitting purposes, we fit the H$\alpha$ complex\footnote{We consider the H$\alpha$ complex as the H$\alpha$ line and the \nii{} doublet surrounding it.} while allowing the global redshift to be a free parameter. We assume an uninformed prior ranging from 0.015 to 0.02 (e.g., \citealt{conselice_nature_2001}; \citealt{gendron-marsolais_revealing_2018}; \citealt{hitomi_collaboration_quiescent_2016}; \citealt{fabian_chandra_2000}); the fit was made using \texttt{emcee}. We report a redshift value of 0.0179 as the optimal value.

\subsection{Chandra X-ray Observatory}\label{sec:methodChandra}

The \textit{Chandra X-ray Observatory} observations presented in this paper were taken with the Advanced CCD Imaging Spectrometer (ACIS) using the \texttt{FAINT} mode.  
We use the following ObsIDs: 3209, 4289, 4946, 4947, 6139, 6145, 4948, 4949, 6146, 4950, 4951, 4953, 11713, 12025, 12036, 12037, 11714. The total combined exposure time is 1.16 Ms.
%In appendix \ref{app:ChandraObservations}, we enumerate the properties of each observation.
% LIST Observations
Each observation was cleaned using a standard methodology implemented using the \texttt{CIAO} (\texttt{v.4.14}) software. The level-one event files were first processed using \texttt{lc\_sigma\_clip} to remove any background X-ray flare events. Then the data were reprocessed using the \texttt{chandra\_repro} command with \texttt{vfaint=true} to take into consideration the diffusion nature of the emission. Background images were constructed using the \texttt{blanksky} routine. Finally, we created a background-subtracted, exposure-corrected, merged image using \texttt{merge\_obs} on the soft band (0.5-2.0 kev). The details of the cleaning process can be found in \cite{rhea_x-tra_2020}. Each pixel in the resulting combined image is approximately 176.5 pc $\times$ 176.5 pc, corresponding to a single instrument pixel ($0.496\times0.496^{"}$) of the Chandra ACIS detectors.

\begin{figure*}[t!]
    \centering
    \gridline{
            \fig{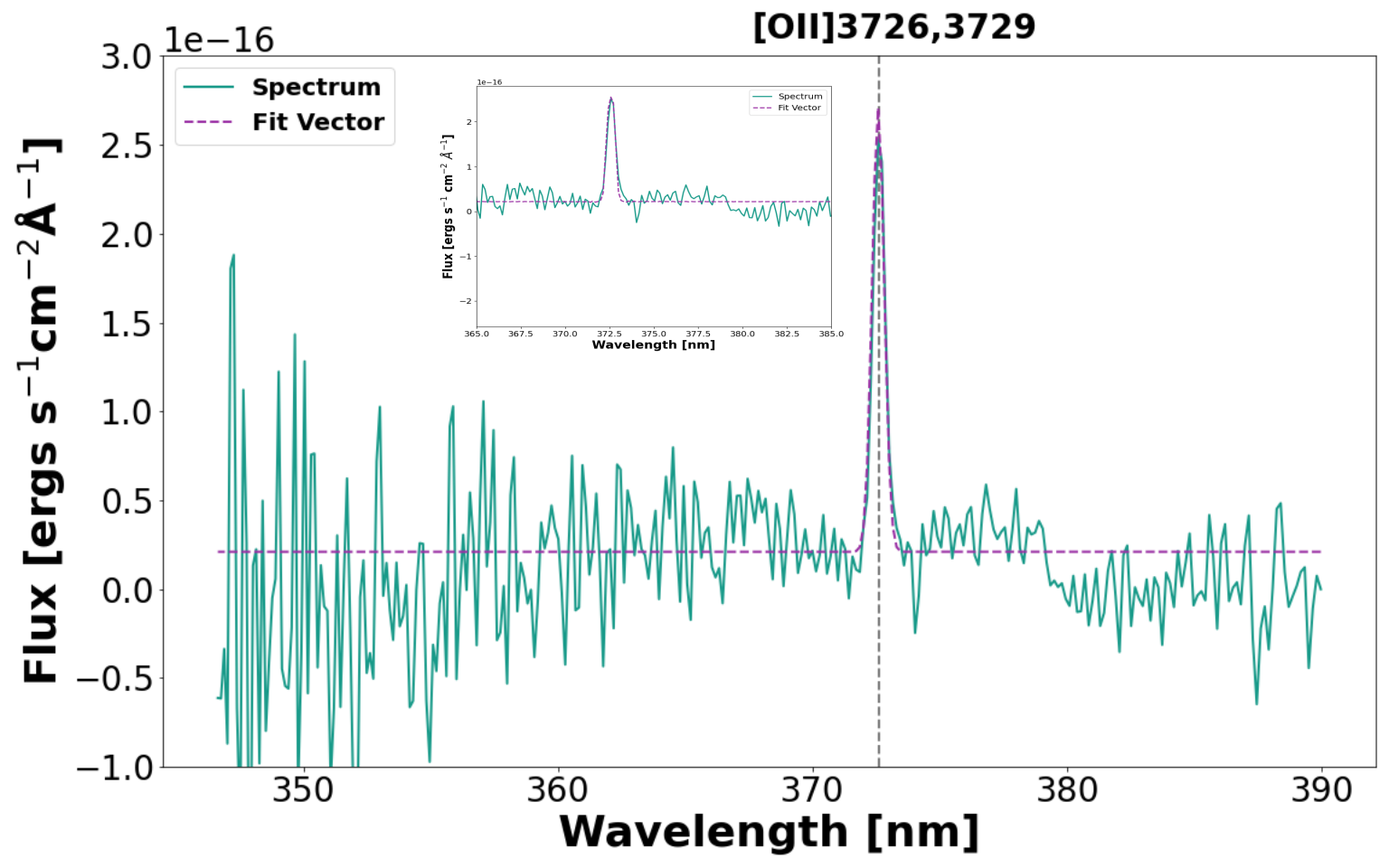}{0.32\textwidth}{(a) SN1}
            \fig{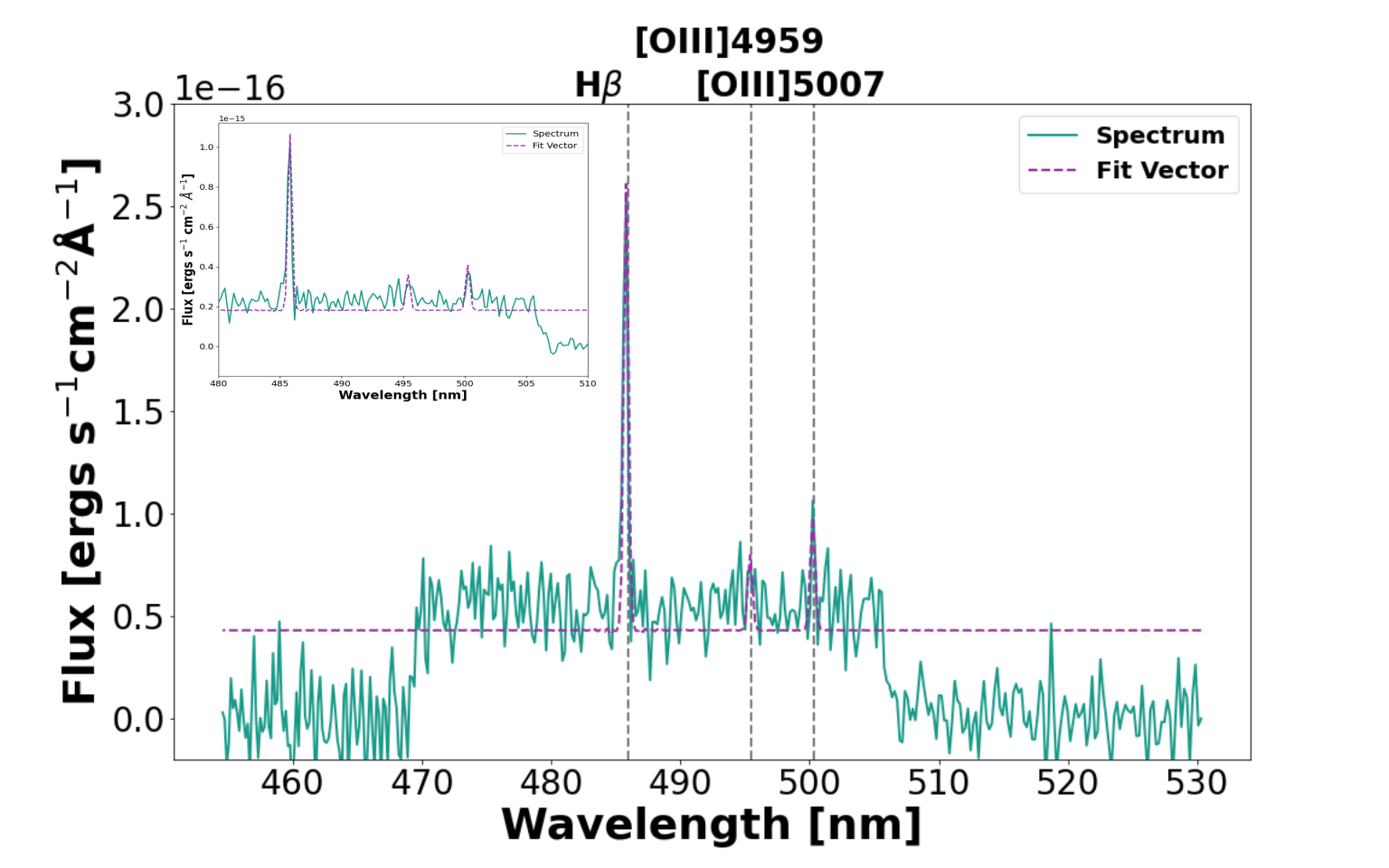}{0.32\textwidth}{(b) SN2}
            \fig{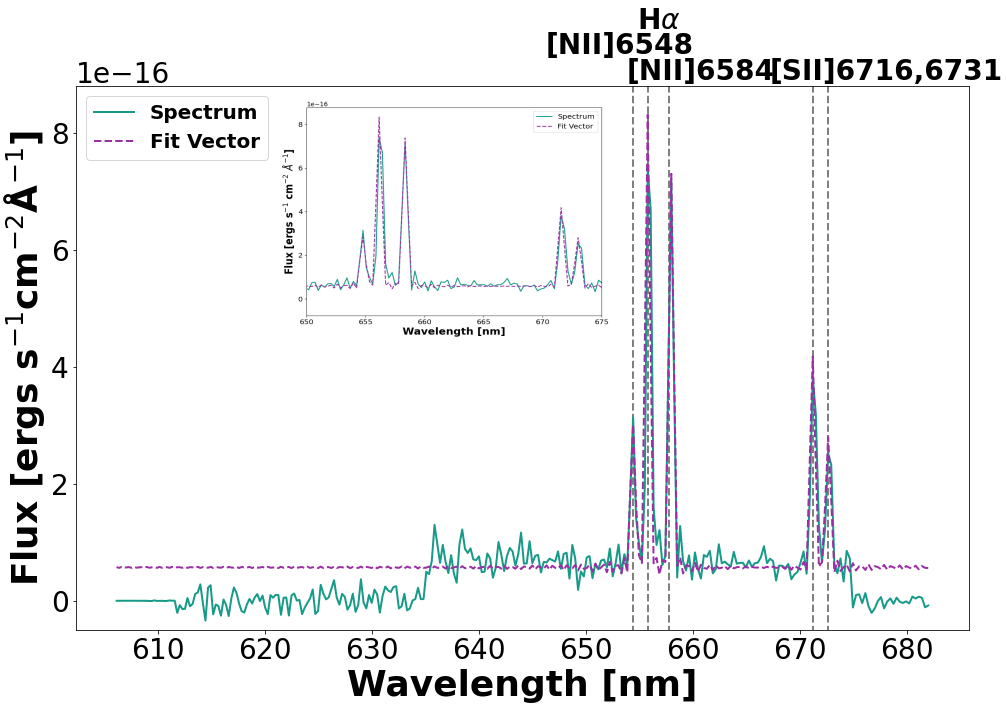}{0.32\textwidth}{(c) SN3}
            }
        \gridline{
            \fig{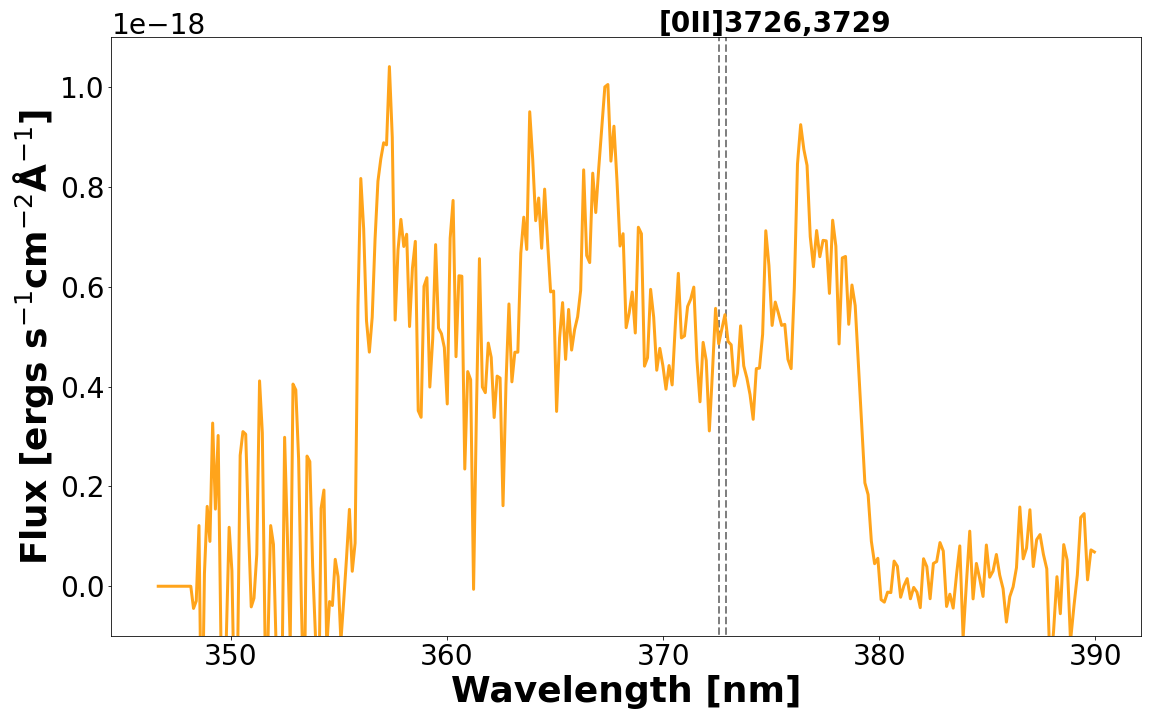}{0.32\textwidth}{(d) SN1 sky emission}
            \fig{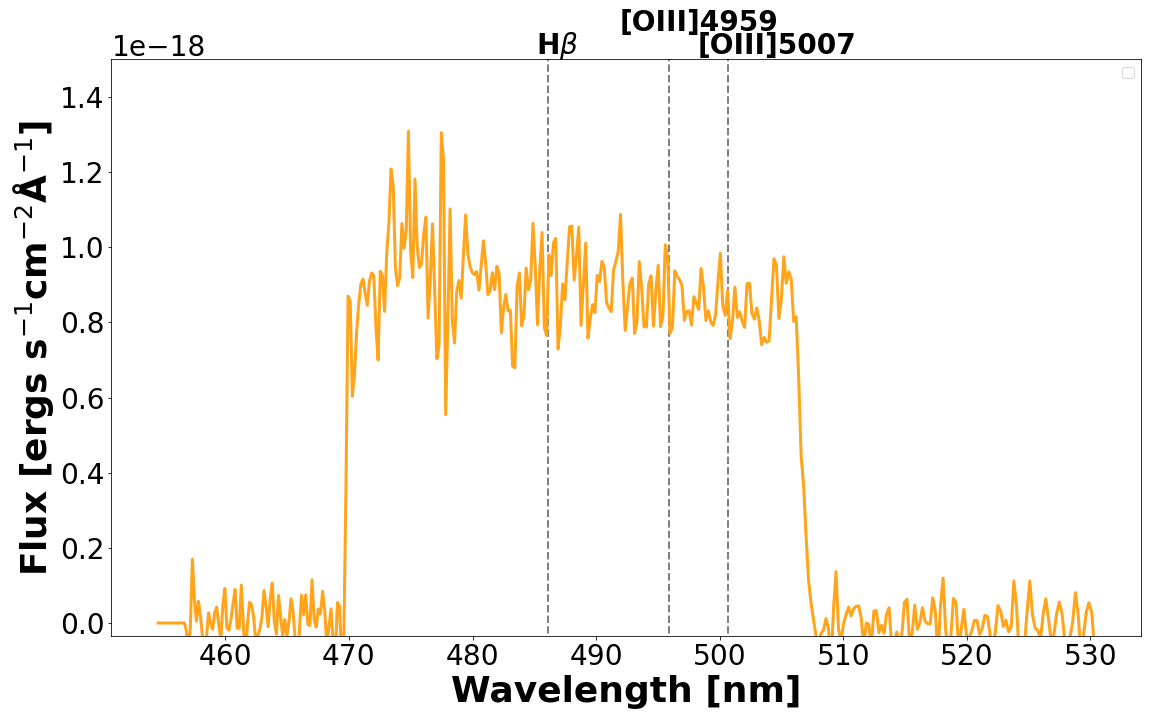}{0.32\textwidth}{(e) SN2 sky emission}
            \fig{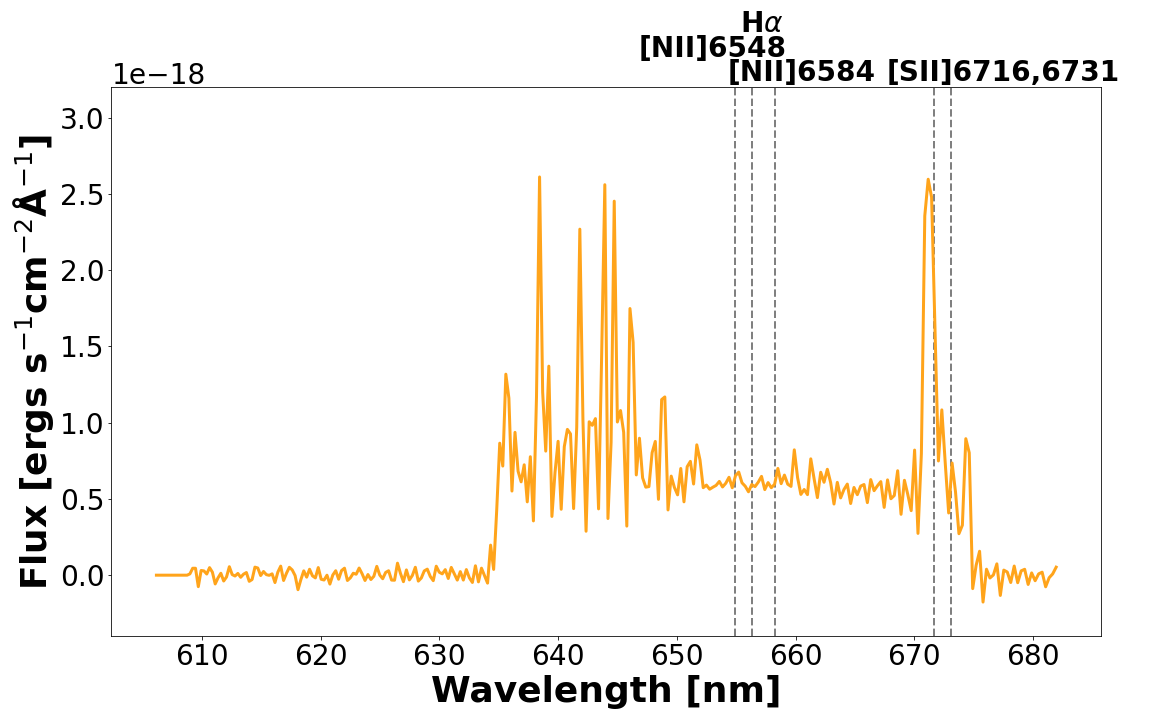}{0.32\textwidth}{(f) SN3 sky emission}
            }
    \caption{Illustrative SN1 (a), SN2 (b), and SN3 (c) spectra (blue) and corresponding fit (purple). The spectrum is integrated from a square region containing 9 pixels (3x3 bin) centered at (\textit{3:19:49.7, +41:30:45.4}) corresponding to the left-hand side of the central region. The x-axis has been shifted by the redshift of NGC 1275. The bottom row shows the SN1 (d), SN2 (e), and SN3 (f) sky emission spectra (orange). Here, the spectrum is integrated from a circular region centered at (\textit{3:19:42.4979, +41:32:01.810}) with a radius of 9.6 arcseconds.}
    \label{fig:illustritiveFits}
\end{figure*}

\section{Results} \label{sec:results}

\subsection{Emission-Line Fits}

In figure \ref{fig:illustritiveFits}, we show the spectrum and the fit for a 3x3 binned spaxel in the central region that contains detectable emissions in SN1, SN2, and SN3; additionally, we show the sky spectrum for each filter. 

The fit of SN1 (panel a) demonstrates that the \oii{}-doublet is in fact present and the two components are indistinguishable. Similarly, the SN2 fit (panel b) shows the strong presence of H$\beta$, the weak presence of \oiii{}$\lambda$5007, and the non-detection of \oiii{}$\lambda$4959 as previously reported in \citet{hatch_origin_2006}. Panel c showing the SN3 fit marks the presence of the strong H$\alpha$ and \nii{}$\lambda$6583 and relatively strong \sii{}-doublet.

\begin{figure*}
    \centering
    \gridline{ 
            \fig{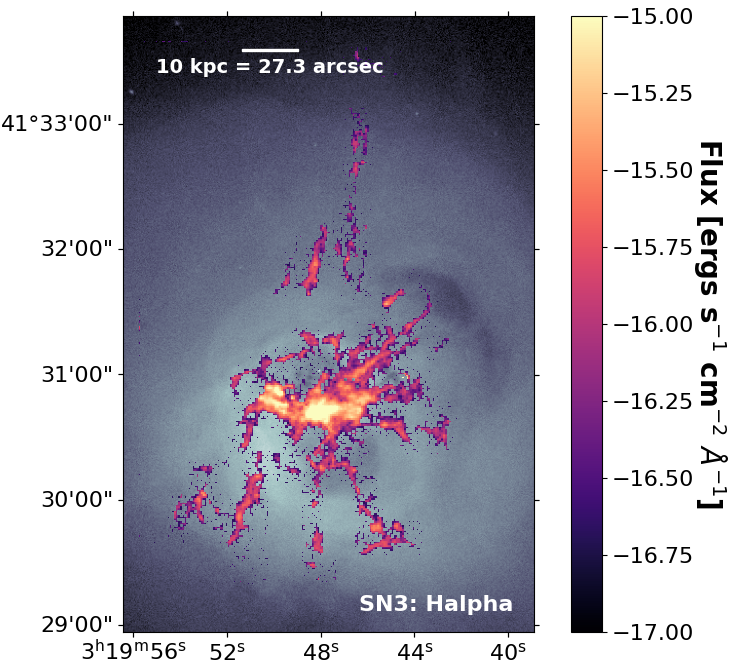}{0.4\textwidth}{}
            \fig{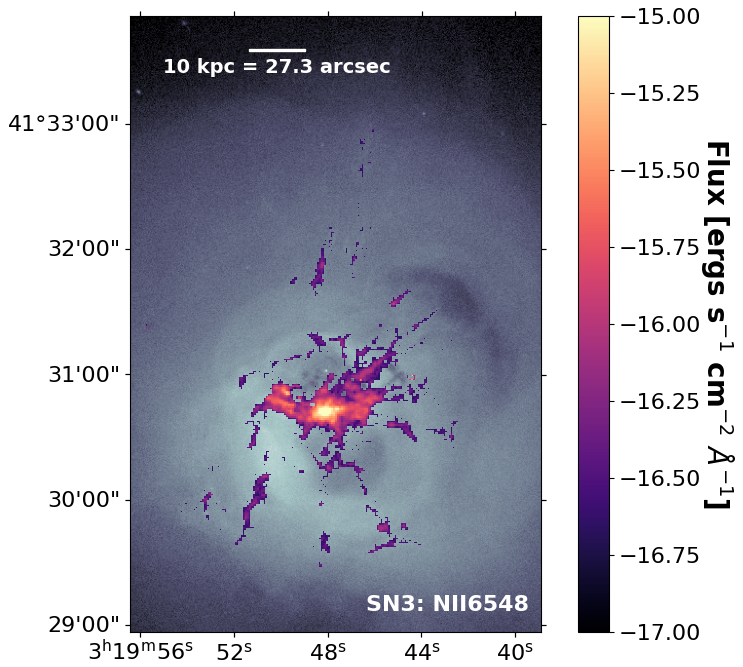}{0.4\textwidth}{}}\vspace{-0.8cm}
    \gridline{ 
            \fig{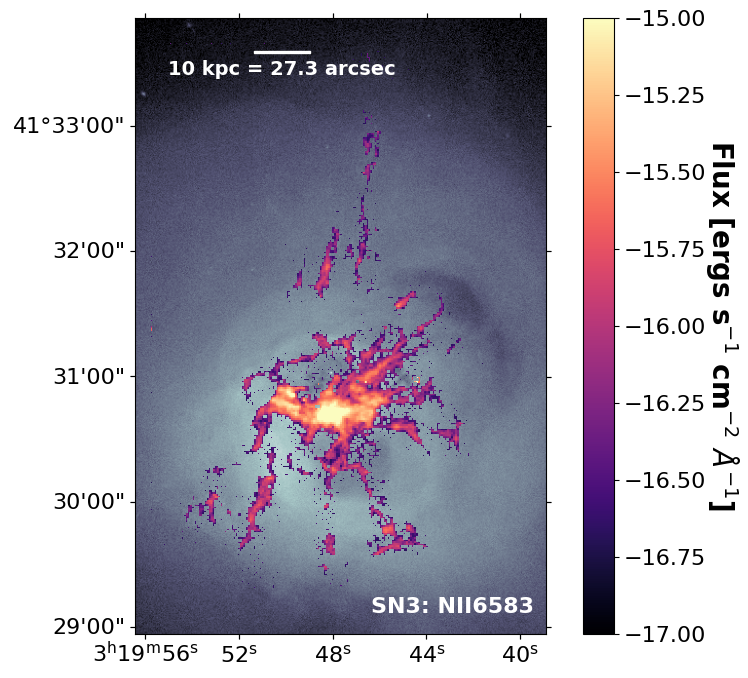}{0.4\textwidth}{}
            \fig{ 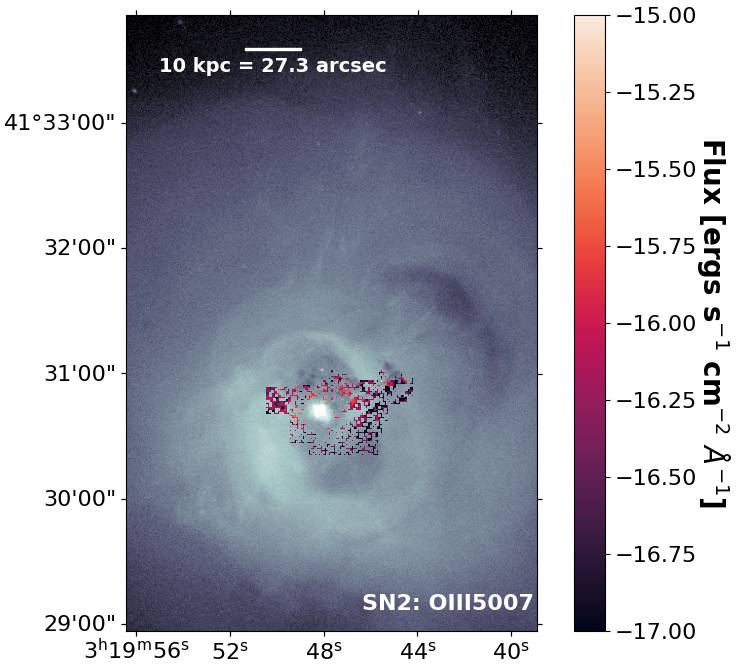}{0.4\textwidth}{}\vspace{-0.8cm}
            }
   \gridline{
            \fig{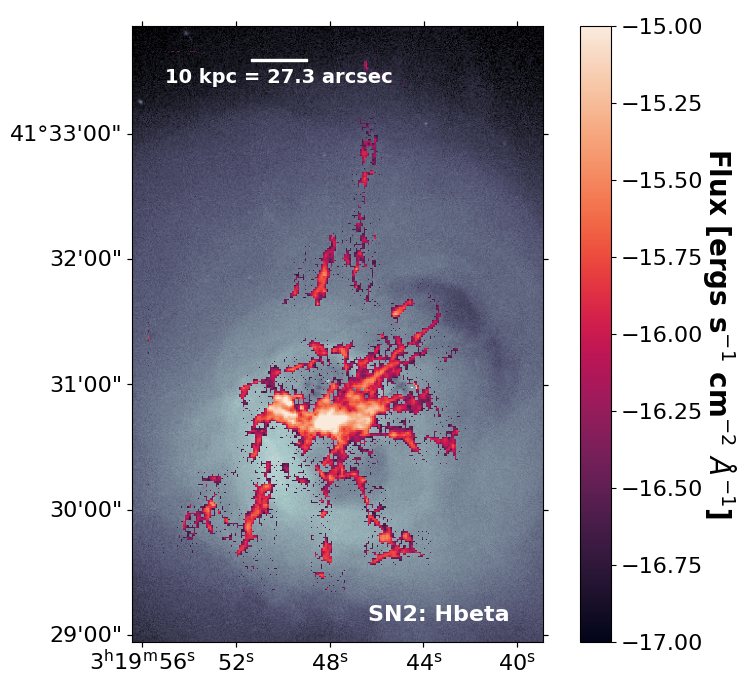}{0.4\textwidth}{}
             \fig{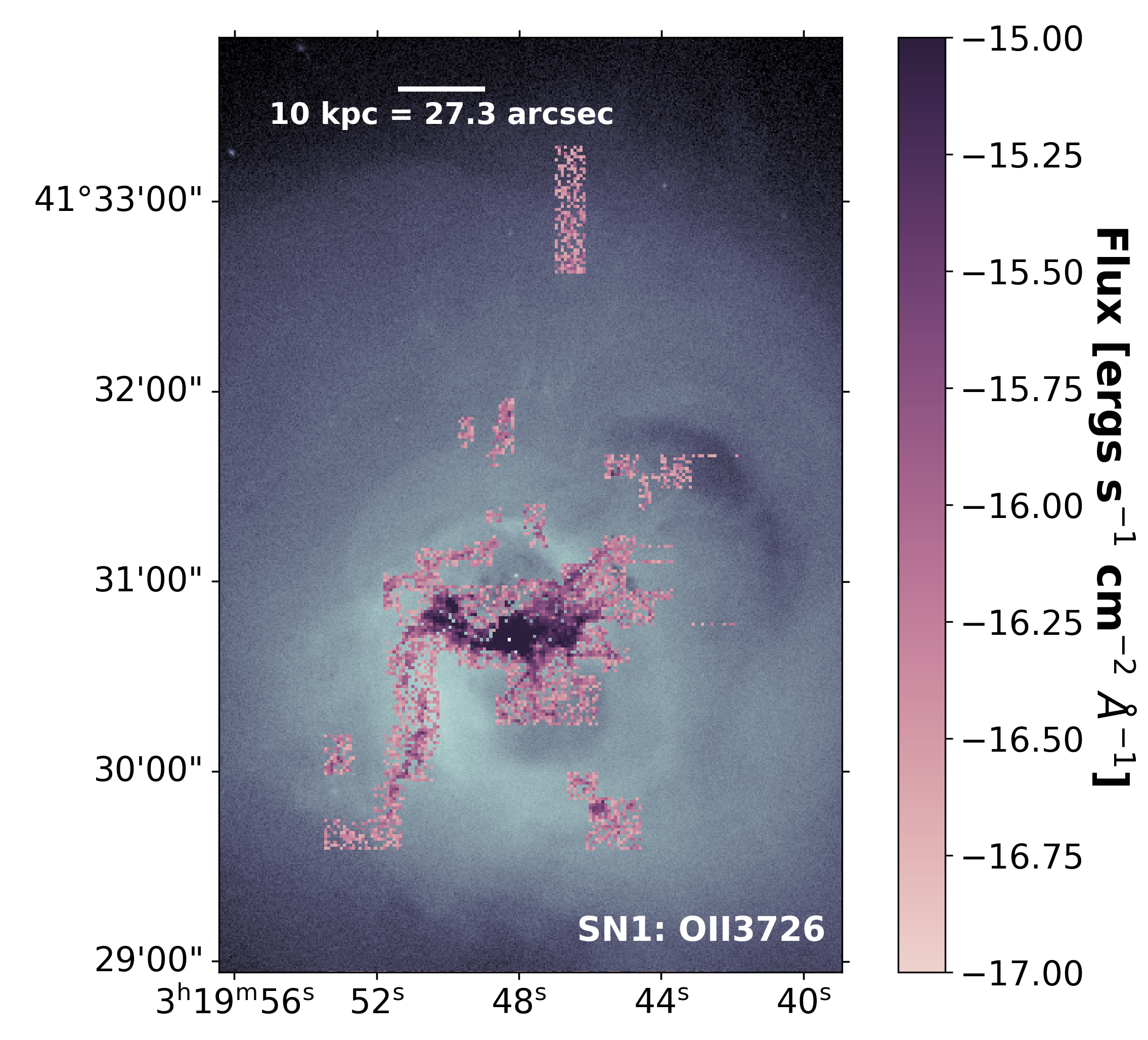}{0.4\textwidth}{}\vspace{-0.5cm}
            }
    \caption{Masked flux maps for each emission line in SN3, SN2, and SN1: H$\alpha$, \nii{}$\lambda$6548, \nii{}$\lambda$6583, \oiii{}$\lambda$5007, H$\beta$,  \oii{}$\lambda$3726+\oii{}$\lambda$3729. The calculated flux maps are masked below a SNR of 3 and a flux cut at 1$\times10^{-17}$ ergs s$^{-1}$ cm$^{-2}$ \AA$^{-1}$. Remaining noisy pixels were removed manually using \texttt{ds9}. In the background, we display the exposure-corrected, background-subtracted, merged \textit{Chandra} image. The images are aligned to approximately 0.7 arcsecs.}
    \label{fig:fluxMaps}
\end{figure*}

Figure \ref{fig:fluxMaps} shows the flux map for each emission line. Similar to the results presented in \cite{gendron-marsolais_revealing_2018}, both the H$\alpha$ and \nii{}$\lambda$6583 emission are omnipresent throughout the nebula in the form of extended filamentary structures reaching dozens of kiloparsecs from the central region. In this work, we did not mask the AGN and can report the strong presence of the H$\alpha$ and the \nii{}$\lambda$6583 throughout the central regions. 

With the SITELLE data, we can mostly separate the HVS from the emission of NGC 1275, because they have distinct velocities (e.g. \citealt{conselice_nature_2001}).
Despite this, a peculiar feature in the flux maps is the lack of nebular emission (from the LVS associated with NGC 1275) in regions overlapping with the HVS. We compared these regions with HST imaging (i.e. \citealt{conselice_nature_2001}) and found that the regions lacking nebular emission were co-spatial with the dust lanes of the HVS as seen through HST imaging. Hence, it appears that the dust in the HVS absorbs significantly the nebular emission originating from NGC 1275, essentially obscuring it from SITELLE.

%The SN2 data, except for the most central regions, suffers from low SNR throughout the nebula. 
H$\beta$ is weakly detected in the nebula and is primarily concentrated in the central region surrounding NGC 1275. In figure \ref{fig:fluxMaps}, we use the H$\alpha$ map to apply a filter on the H$\beta$ map; we do this only after verifying that the H$\beta$ emission follows the same contours. This process is done to reduce the amount of noise in the image.
%Importantly, we detect weak \oiii{} emission at $\lambda$5007 but not its counterpart at $\lambda$4959. This result is consistent with previous findings (i.e. \citealt{hatch_origin_2006}).
%Despite the fact that it is weakly detected throughout the nebula, the \oiii{} emitting gas does not appear to be smoothly dispersed but rather knotty in appearance (see figure \ref{fig:fluxMaps}h).
Moreover, \oiii{}$\lambda$5007 is predominant in the central regions but missing in the extended filaments.  We discuss this further in Section \ref{sec:discussion}. %Therefore, we do not associate the \oiii{} emission with the nebula but rather the underlying stellar population.

Since the SN1 observation does not have sufficient spectral resolution to disentangle the two components of the \oii{}-doublet, we combine their calculated fluxes (see figure \ref{fig:fluxMaps}f). The flux map reveals the omnipresence of this strong emission-line throughout the nebula. However, due to large errors in the flux owing to the low signal-to-noise, we are unable to draw conclusive results from trends in the \oii{} emission.

%\begin{figure*}
%    \centering
%    \gridline{
%            \fig{Background_SN1.png}{0.32\textwidth}{(a) SN1 Background}
%            \fig{Background_SN2.png}{0.32\textwidth}{(b) SN2 Background}
%            \fig{Background_SN3.png}{0.32\textwidth}{(b) SN3 Background}
%            }
%    \caption{SN1 (a), SN2 (b), and SN3 (c) background spectra (orange). The spectrum is integrated from a circular region centered at (\textit{3:19:42.4979, +41:32:01.810}) with a radius of 9.6 arcseconds.}
%    \label{fig:backgrounds}
%\end{figure*}

\subsection{Emission Line Classification}

Emission line ratios and equivalent widths  have been used to classify emission line nebulae. However, emission-line classification classification systems (BPT and WHAN) based on star-forming galaxies with and without AGN may fall short of diagnostic for emission line regions with more physical processes in play. We show that traditional classification systems may be useful in identifying properties that are similar to those observed in emission-line galaxies forming stars, but have limited diagnostic usefulness beyond that.  We use traditional BPT diagram diagrams to  plot emission line ratios from the central portion of the nebula where \oiii{} is detected, and we find that the emission line ratios in the central region are inconsistent with AGN photoionization or star formation as a sole source of energy and ionization. We show a weak trend of \oiii{}/H$\alpha$ with distance from the AGN which indicates that AGN photoionization may contributing to the ionization state of the gas very close to the AGN. But the constant \nii{}/H$\alpha$ ratio through the same region suggests that the dominant source of energy is more distributed. We use the WHAN diagram to inspect and classify \nii{}/H$\alpha$ emission line ratios and Ha equivalent widths. Inspection of the WHAN diagrams for the line emission from the even more extended nebular gas is shows emisison with high equivalent width and \nii{}/H$\alpha$ ratios not that dissimilar from what is observed in the center. The range is broader, with a hint of an inverse relationship between equivalent width of H$\alpha$ and the \nii{}/H$\alpha$ ratio.   We discuss these results in turn below.

\subsubsection{Standard BPT Diagram}
%\carter{Describe the \oiii{}/H$\beta$ vs \nii{}/H$\alpha$ diagram process and how we can interpret it.}
BPT diagrams are frequently used to study the ionization mechanism at play in emission-line nebula (e.g., \citealt{baldwin_classification_1981}). The two standard BPT diagrams plot either \nii{}$\lambda$6583/H$\alpha$ or (\sii{}$\lambda$6716+\sii{}$\lambda$6731)/H$\alpha$ versus \oiii{}$\lambda$5007/H$\beta$. %Since the contamination of skylines complicates an accurate measurement of the \sii{}-doublet, 
Since the edge of the filter and residual from the strong sky lines affect the quality of the fit of the \sii{}-doublet, we limit our analysis to the \nii{}$\lambda$6583/H$\alpha$ versus \oiii{}$\lambda$5007/H$\beta$ BPT diagram. 

%BPT diagrams are generally paired with diagnostic cuts that break areas of the diagrams ionized by different mechanisms, from the star-forming area is ionized by massive stars, the Seyfert-like area is ionized by an AGN, and LINER\footnote{Low-ionization nuclear emission-line region}-like area is ionized by an ensemble of sources linked to the nuclear emission of the galaxy: \citealt{kauffmann_host_2003} (K03),  \citealt{kewley_host_2006} (K06), and \citealt{stasinska_semi-empirical_2006} (S06).

While plotting emission line ratios on pre-calibarated regions of BPT diagrams is traditionally done to attempt to differentiate star-forming (SF) regions from regions dominated by AGN, e.g. Seyferts and LINERS (\citealt{kauffmann_host_2003} (K03),  \citealt{kewley_host_2006} (K06), and \citealt{stasinska_semi-empirical_2006} (S06)), the literal classifications are only
useful for emission line nebulae where there are only two options for energy sources: AGN and stars. Figure 4 shows that even in the central portion of NGC1275, all of the points lie in the “Composite” region that suggests that even in the context of only two energetic candidates,  some combination is needed. Therefore, we adopt a more data- and physics-centered approach.

%Traditionally, points lying within the star-forming region are interpreted as necessarily star-forming, the points in the Seyfert/AGN region are interpreted as belonging to an AGN, and point in the LINER region are interpreted as belonging to nuclear emission region. 

However, these lines do not represent steadfast delineations between different types of ionization process. Instead, we adopt a more modern interpretation of the diagnostic lines that states regions in the star-forming area of the diagram must not have a highly energetic ionizing source while regions in either the strong AGN or weak AGN regions must include a highly energetic ionizing source (e.g. \citealt{stasinska_semi-empirical_2006}; \citealt{garn_predicting_2010}; \citealt{cid_fernandes_comprehensive_2011}; \citealt{curti_what_2022}).
Emission line ratios indicating high \oiii{}/H$\alpha$ indicate a high photoionization parameter, or a high ratio of ionizing photons to ambient gas density. A high \nii{} ratio, on the other hand, for gas with low to no \oiii{}, is a measure of the energy per ionization. \nii{} is a significant radiative coolant of ionized gas, and its emission strength is more sensitive to the temperature of the gas than it is to nitrogen abundance.
Additionally, the regions in the Seyfert/AGN area have a stronger highly energetic ionizing source than the regions in the LINER area.
Moreover, it is possible, and highly likely, that a combination of softer (i.e. photoionization) and highly energetic (i.e. cosmic rays or AGN activity) ionizing sources are required to ionize the bulk of the gas. Therefore, the position of the spaxel in the BPT diagram is simply indicative of the relative contribution of soft and highly energetic ionizing photons to the emission, if shocks and particle heating are not considered. It is important to note that high-energy photons have been demonstrated to reproduce emission line ratios in the AGN region of the BPT diagram using CLOUDY simulations; we emphasize this point to indicate that a spaxel lying in the region traditionally denoted as AGN does not mean that an AGN is responsible for the ionization in this pixel (e.g., \citealt{donahue_photoionization_1991}).

\begin{figure}
    \centering
    \includegraphics[width=0.495\textwidth]{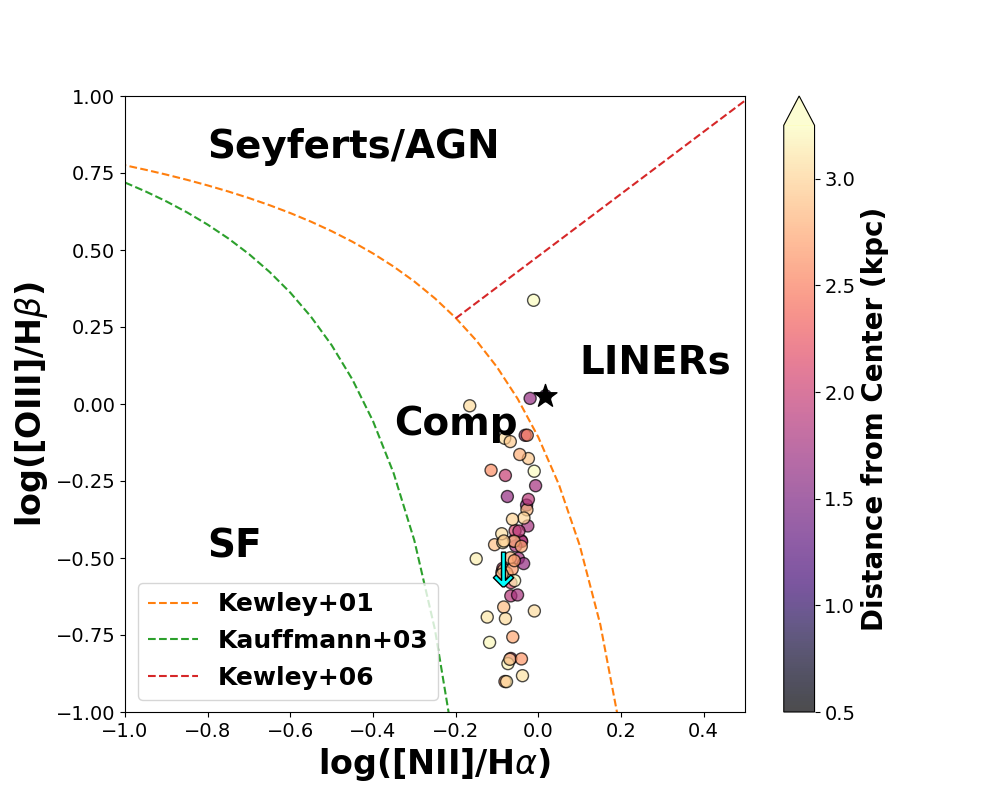}
    \caption{BPT diagram showing log(\nii{}$\lambda$6583/H$\alpha$ versus log(\oiii{}5007/H$\beta$) in the central region of NGC 1275. The points are color coded by their distance to the AGN at the center of NGC 1275. The cyan arrow represents the upper bound the brightest pixel in the small northern filament. The AGN is marked by a black star.}
    We overplot the \cite{kewley_optical_2001} (purple), \cite{kauffmann_host_2003} (orange), and \cite{kewley_host_2006} (blue) diagnostic lines. The region between the \cite{kewley_optical_2001} and the \cite{kauffmann_host_2003} lines is considered a composite region. Spaxels falling in the composite region are considered to be ionized by multiple types of ionizing sources with a range of characteristic average energy per ionization particle or photon.
    \label{fig:BPT}
\end{figure}

Figure \ref{fig:BPT} shows the \nii{}$\lambda$6583/H$\alpha$ versus \oiii{}$\lambda$5007/H$\beta$ BPT diagram. Since \oiii{}$\lambda$5007 is only present in the central region of NGC 1275, the figure is only for the central region with the exception of the blue arrow.
We measure the 3-$\sigma$ detection limit of \oiii{}$\lambda$5007 for the brightest pixel in the small northern filament and include this point as a cyan arrow. The detection limit is calculated by taking a region of the spectrum not containing any lines and calculating the square root of the standard deviation of the flux in the region.
Indeed, the trend highlighted in figure \ref{fig:BPT} indicates that the \oiii{} emission falls off quickly as a function of the distance from the center. The points are color-coded by their distance in kiloparsec to the AGN. The figure reveals two main takeaways: all points lie above the \citealt{kauffmann_host_2003} line, and there is a clear trend in the position of the points in the diagram as a function of their distance from the AGN. First, it demonstrates that photoionization by AGN or by stars alone is not sufficient to explain the ionization of the gas. This is consistent with previous classifications of NGC 1275 as a Seyfert 1.5 type galaxy (e.g., \citealt{osterbrock_seyfert_1981}). Secondly, it indicates that the strength of the ionizing source decreases as a function of distance from the central AGN. We delve further into these in $\S$ \ref{sec:discussionStandardBPT}

\begin{figure}
    \centering
    \includegraphics[width=0.495\textwidth]{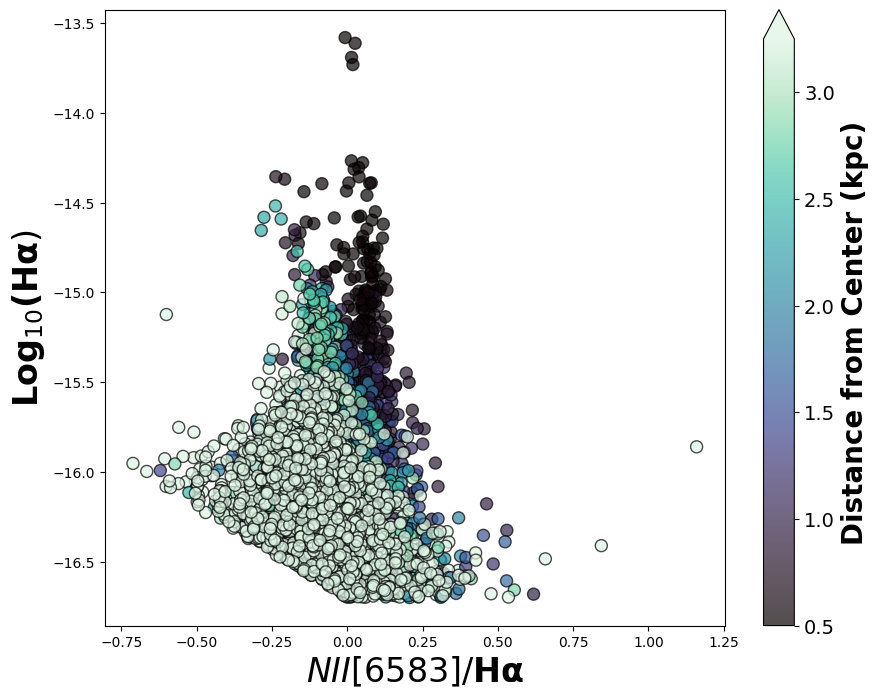}
    \caption{Emission line diagram showing the relationship between log(\nii{}/H$\alpha$) and log(H$\alpha$) as a function of the distance from the AGN (designated here at the center) in kpc. The central 0.5kpc have been excised to remove any issues pertaining to multiple emission lines. 
    }
    \label{fig:NIIHalpha_Halpha}
\end{figure}

In figure \ref{fig:NIIHalpha_Halpha}, we show the relationship between log(\nii{}/H$\alpha$) and log(H$\alpha$) as a function of the distance from the AGN. The figure shows a distinct trend in the relationship as the spaxels become further from the influence of the central AGN. Spaxels representing points nearest the AGN have consistent \nii{}/H$\alpha$ ratios regardless of intrinsic H$\alpha$ brightness while points further from the AGN cover a wide range of ratios. This plot exposes a distinct population of points that are both bright and a considerable distance from the AGN (at least 2 kpc) with a characteristic \nii{}/H$\alpha$ near 0. The population of distant filaments (more than 3 kpc from the center) with lower surface brightness do not have a single characteristic \nii{}/H$\alpha$ ratio but rather have a noteable spread in values.

\subsubsection{WHAN Diagram}\label{sec:resultsWHAN}

Since the \oiii{}$\lambda$5007 emission in the extended nebula is too faint to be detected, we cannot rely on the standard BPT diagram to give a complete picture of the ionization mechanism in NGC 1275; therefore, we rely on the so-called WHAN diagram (e.g. \citealt{cid_fernandes_alternative_2010}; \citealt{cid_fernandes_comprehensive_2011}; the name WHAN comes from equivalent Width H$\alpha$-Nitrogen). This diagnostic plot was conceived to investigate the main ionization mechanism in weak line galaxies (WLGs). WLGs are defined as galaxies for which a strong measurement (SNR$>$3) of \oiii{}$\lambda$5007 and/or H$\beta$ is impossible. Therefore, this plot serves as an appropriate choice for NGC 1275.

Although we refer the reader to \cite{cid_fernandes_alternative_2010} and \cite{cid_fernandes_comprehensive_2011} for a comprehensive dive into this diagram and how it compares with the standard BPT diagram, we recast the discussion on the discriminating lines in the WHAN diagram here since they are notoriously difficult to interpret. \textbf{We again stress the need to not overinterpret these lines} (e.g. \citealt{stasinska_semi-empirical_2006}; \citealt{cid_fernandes_comprehensive_2011}; \citealt{curti_what_2022}; \citealt{garn_predicting_2010}). More precisely, the \nii{}/H$\alpha$ diagnostic cuts do not delineate between stars and AGN but rather between pixels that are consistent with stellar photoionization and pixels where photoionization is unlikely to be the dominant source of energy.

Similar to standard the BPT diagrams, there are three separate diagnostic lines used to distinguish spaxels in which star-formation (or the presence of massive O and/or B stars) can explain the measured line ratio as opposed to requiring a highly energetic ionizing source such as an AGN: KO3, KO6, and SO6. Following the reasoning described in \cite{cid_fernandes_comprehensive_2011}, we adopt the S06 line cut at log(\nii{}$\lambda$6583/H$\alpha$)=-0.4 such that any point to the left of this line falls in the star-forming category and any points to the right fall either in the Seyfert or LINER classification. Additionally, this line is not a hard cut-off, meaning not all spaxels to the left are star-forming, and all to the right are not star forming (see \citealt{stasinska_semi-empirical_2006}, \citealt{cid_fernandes_alternative_2010}, and \citealt{cid_fernandes_comprehensive_2011} for details). Rather, spaxels to the left of this line do not require a hard ionizing source (HIS) to explain the line ratio. Similarly, we use the transposed K06 Seyfert/LINER diagnostic line which lies at W$_{H\alpha}$=6\AA,
where W$_{H\alpha}$ is the equivalent width and is defined as the H$\alpha$ flux over the continuum flux (e.g., \citealt{valeasari_less_2020}).
Any spaxel above this line is generally classified as a HEW and any spaxel below this line is classified as a LEW. The distinction here comes from the fact that the equivalent width measures the ratio of the contributions of the highly energetic ionizer and the non-ionizing stellar component; therefore, the sHIS classification refers to a strong hard-ionizing agent while a wHIS refers to a weaker hard-ionizing agent. We do not use the standard nomenclature of strong AGN and weak AGN since, in the Perseus cluster, literature suggests that the AGN does not play a role in the ionization of the outer filaments (e.g., \citealt{johnstone_extended_1988}). Rather, these regions are likely ionized by some other physical phenomena  exhibiting a power-law spectrum such as from the cooling of hot X-ray gas or magnetic reconnection (e.g., \citealt{ferland_collisional_2009}; \citealt{werner_origin_2014}; \citealt{fabian_energy_2011}). We discuss the specifics of these mechanisms in section 4.2.4. Instead, we use data-driven names such as high equivalent width (HEW) and low equivalent width (LEW).
Comparatively, the SF region does not require a hard-ionizing agent to explain the line ratio.  There is an additional diagonal, dotted line below which measurements are uncertain (\citealt{cid_fernandes_comprehensive_2011}).

Figure \ref{fig:WHANbyRegion} presents the WHAN diagram for NGC 1275. We show the position of each pixel in the WHAN plot color-coded by the region to which they belong. Moreover, we overplot the mean value for each region and the associated 1-$\sigma$ error. The figure reveals that most points lie within the Seyfert designation. This implies that a relatively-strong highly energetic ionizing source is required to ionize the gas in these spaxels. On the right-hand side of the image, we present the spatial distribution of pixels color-coded by their WHAN plot designation. We note that the vast majority of the nebula has a highly energetic ionizing agent (thus classified as Seyfert) while some boundaries of the inner filaments are characterized as LEW,  implying that some boundaries seem to require a weaker hard-ionizing agent. We note, however, that the distinction of weak versus hard only stands for photon-based ionization and does not cover the scenario in which the ionizing source is from a different type of particle. Finally, the spaxels designated as SF are primarily contained within the blue-loop (\citealt{canning_star_2010}). We discuss each region of the nebula in more detail in $\S$\ref{sec:regionDiscussion}.

\begin{figure*}
    \gridline{
        \fig{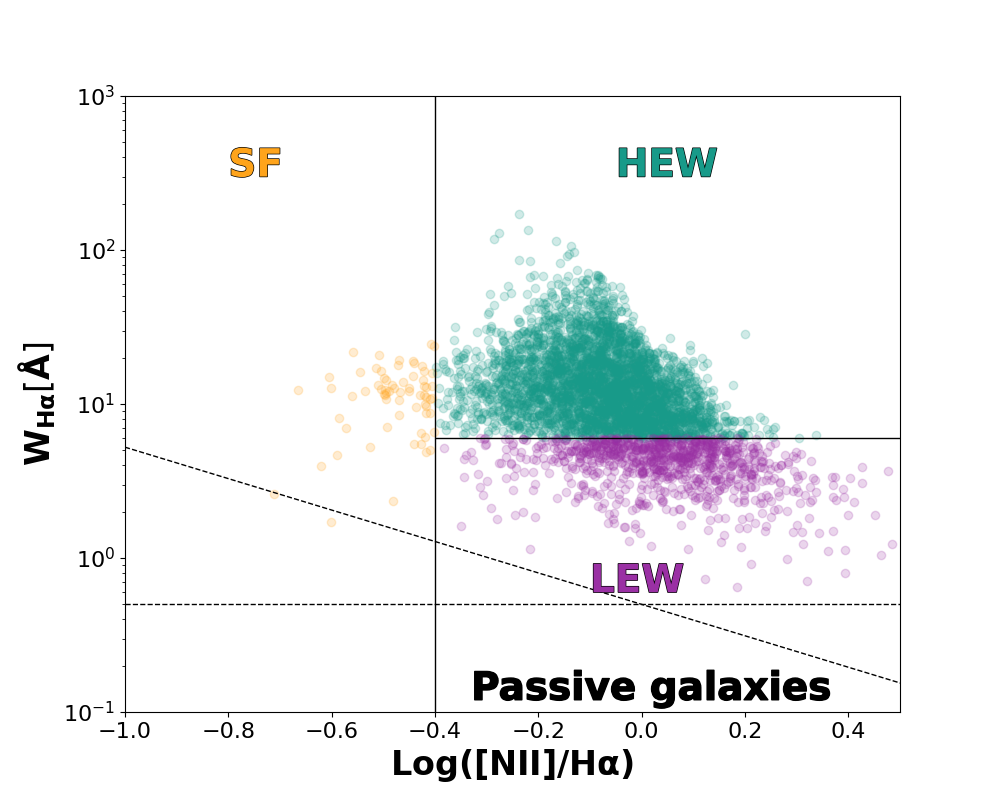}{0.57\textwidth}{(a) WHAN diagram of emission pixels.}
        \fig{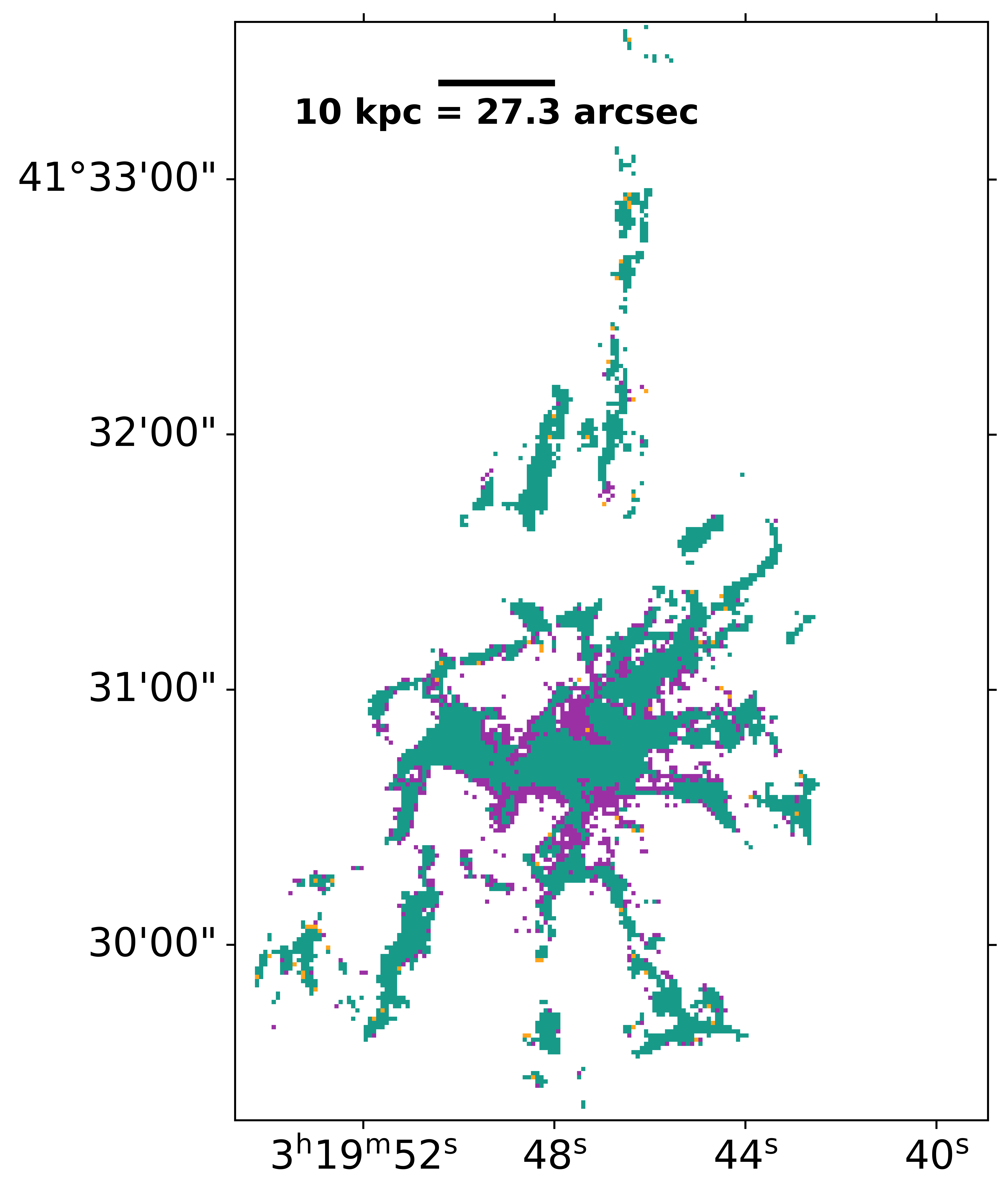}{0.37\textwidth}{(b) Color-coded emission-line nebula map.}
    }
    \caption{WHAN diagram and color-coded map for all pixels in the nebula surrounding NGC 1275 with SNR over 3 and flux over 1e$^{-17}$ ergs s$^{-1}$ cm$^{-2}$ \AA$^{-1}$. The pixels are divided into 3 categories: star-forming (orange), HEW (teal), and LEW (purple). Measurements below the dotted line are considered uncertain \citep{cid_fernandes_comprehensive_2011}.}
    \label{fig:colorcoded}
\end{figure*}

\begin{figure*}
    \gridline{
        \fig{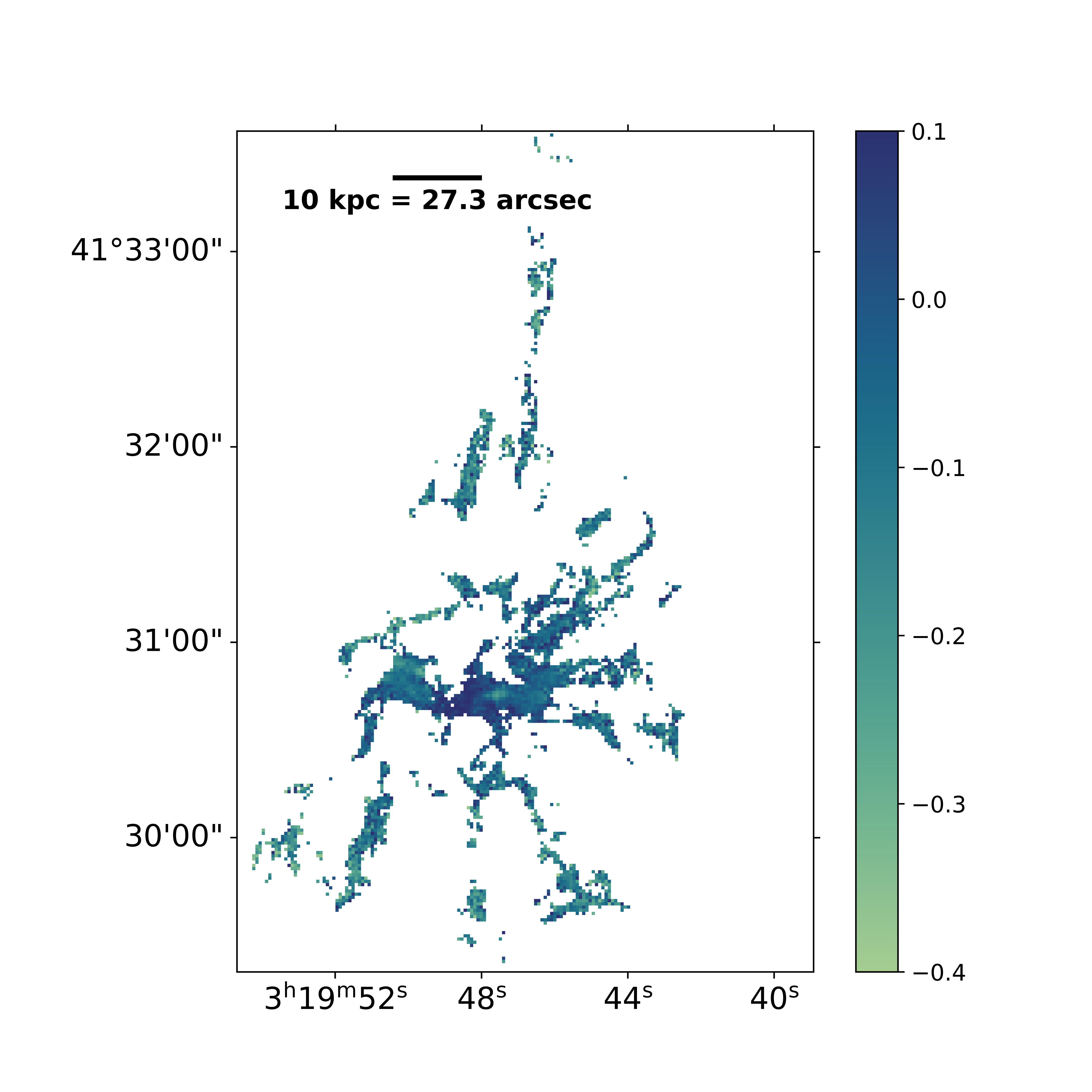}{0.47\textwidth}{(a) Log(\nii{}/H$\alpha$) values for HEW spaxels.}
        \fig{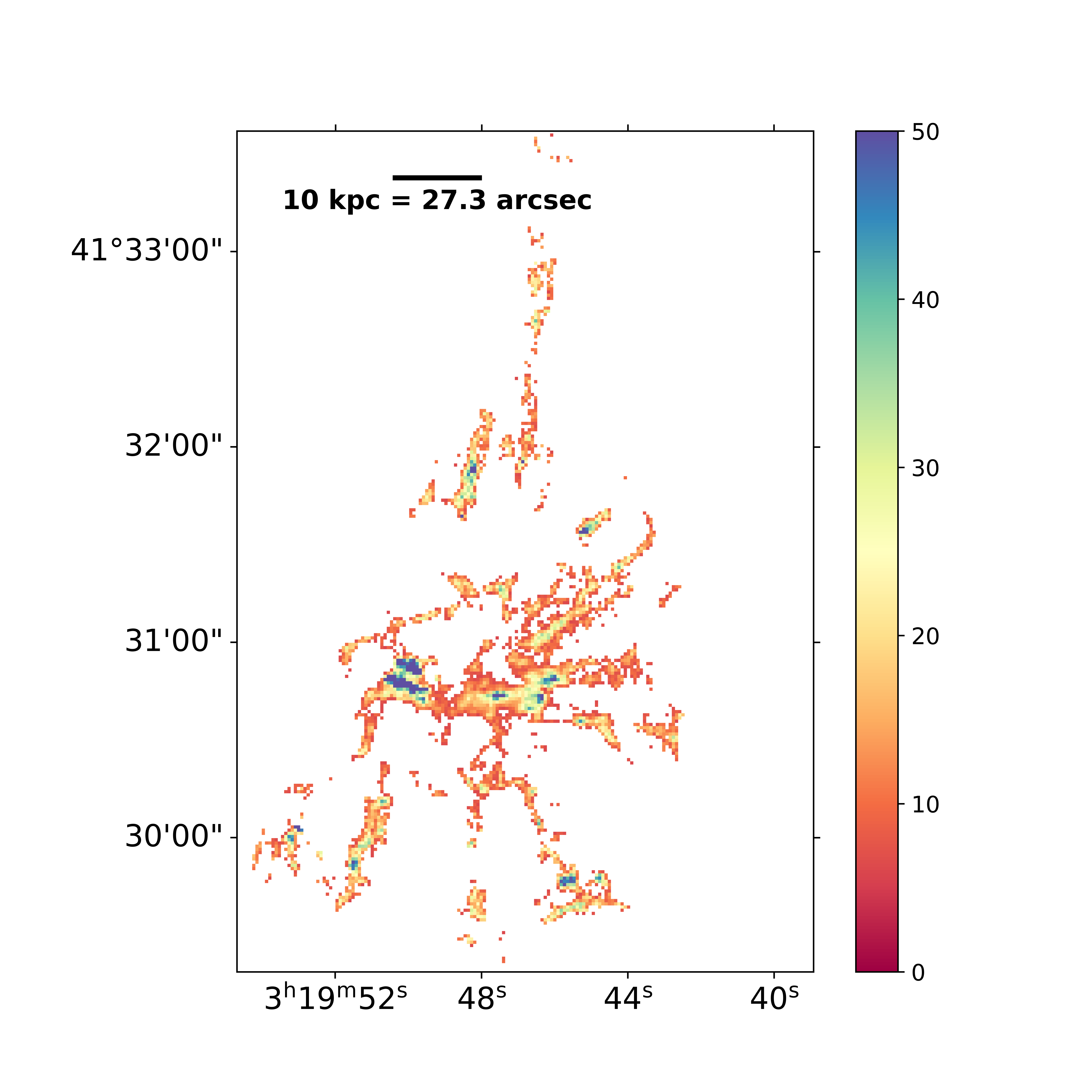}{0.47\textwidth}{(b) Equivalent width values for HEW spaxels in \AA.}
    }
    \caption{In the left-hand diagram, we plot the log(\nii{}/H$\alpha$) values for spaxels labeled as HEW. Similarly, in the right-hand panel, we plot the equivalent width values for spaxels labeled as HEW in units of \AA.}
    \label{fig:colorcoded-seyfert}
\end{figure*}

\section{Discussion} \label{sec:discussion}
In this section, we discuss the results in the context of the standard BPT diagram, the WHAN diagrams, and overall trends.

\subsection{Standard BPT region analysis}\label{sec:discussionStandardBPT}
%\subsubsection{OIII in core}
%Ionization outside of the core of NGC 1275 is not sufficiently hard to doubly ionize oxygen. We obtain clear detection of OII in the filaments -- so the issue isn't lack of oxygen, but instead lack of ionizing factor.

The middle right panel of figure \ref{fig:fluxMaps} shows that the \oiii{}$\lambda5007$ flux in NGC 1275 is well constrained to the central regions of the galaxy, and we do not find it in the extended filaments.
%and does not trace the ionized gas nebula.  
In figure \ref{fig:BPT}, we plot the log(\nii{}$\lambda$6583/H$\alpha$) vs. log(\oiii{}$\lambda5007$/H$\beta$) BPT diagram for the central regions of NGC 1275 color-coded by the distance to the center of the galaxy. There is a clear trend that regions close to the center fall in the HEW and LEW regions and that the log(\oiii{}/H$\beta$) ratio decreases as a function of the distance. Together, this indicates that while the ionizing source remains highly energetic the strength of the source decreases further from the center. This strengthens the argument that the AGN at the center of NGC 1275 plays a role in the ionization of, at least, the central region of the nebula. Moreover, the BPT diagram reveals that photoionization from O or B stars is not sufficient to explain the ionized gas throughout the region where \oiii{} is detected. In the filaments where we do not detect \oiii{}, the upper bound limit shows that the photoionization is again not sufficient to explain the ionization of the filaments. Moreover, when compared with the position of typical galactic regions on the standard BPT plot (see for example \citealt{veilleux_spectral_1987}), the position of this upper bound indicates that the ionization mechanism at play in the outer filaments is distinct from ionized gas regions in our galaxy or nearby AGN.

%\begin{figure}
%    \centering
%    \includegraphics[width=0.49\textwidth]{OIIvsOIII_Map.png}
%    \caption{
%    \oii{}$\lambda$3726+\oii{}$\lambda$3729 versus \oiii{}$\lambda$5007.
%    }
%    \label{fig:OIIvsOIII}
%\end{figure}
%- Trend of highly energetic ionizing source closer to the AGN (figure \ref{fig:BPT}). Similar to \url{https://ui.adsabs.harvard.edu/abs/2017MNRAS.467.3399W/abstract} figure 8.

%- This is expected naively

%Mention main results then go into detail (like Tremblay+)
\subsection{WHAN Plot Analysis}\label{sec:regionDiscussion}
Since the measurable \oiii{} emission is constrained to the core of NGC 1275, we rely on the WHAN diagram described in $\S$\ref{sec:resultsWHAN} to analyze the ionization mechanism at larger scales in the nebula. We are also interested in the trends observed in the main structures previously identified in the nebula (see figure \ref{fig:combinedImage}).

In figure \ref{fig:colorcoded}, we show the WHAN diagram for all the spaxels in SN3 that belong to the nebula and have a SNR greater than 3. On the right-hand side, we show the spaxels color-coded by emission type as defined by their location in the WHAN diagram in Right Ascension vs. Declination. Although the vast majority of the spaxels in the nebula lie within the Seyfert designation in the WHAN plot, there are populations of spaxels residing in the star-forming region and the LEW region. The majority of the star-forming spaxels fall within the blue-loop as defined in \cite{canning_star_2010}. The LEW spaxels are primarily located in the outskirts of the central region. Overall, this indicates that the equivalent width is not an accurate indicator of AGN activity since the outer filaments are not expected to be affected by the AGN.

In the left panel of figure \ref{fig:colorcoded-seyfert}, we plot only spaxels categorized as HEW color-coded by their log(\nii{}/H$\alpha$) value. The log(\nii{}/H$\alpha$) value is a proxy for the strength of the ionizing agent. On the right panel of figure \ref{fig:colorcoded-seyfert}, we again plot the spaxels categorized as HEW but this time they are color coded by W$_{H\alpha}$.

\begin{figure}
    \centering
    \includegraphics[width=0.495\textwidth]{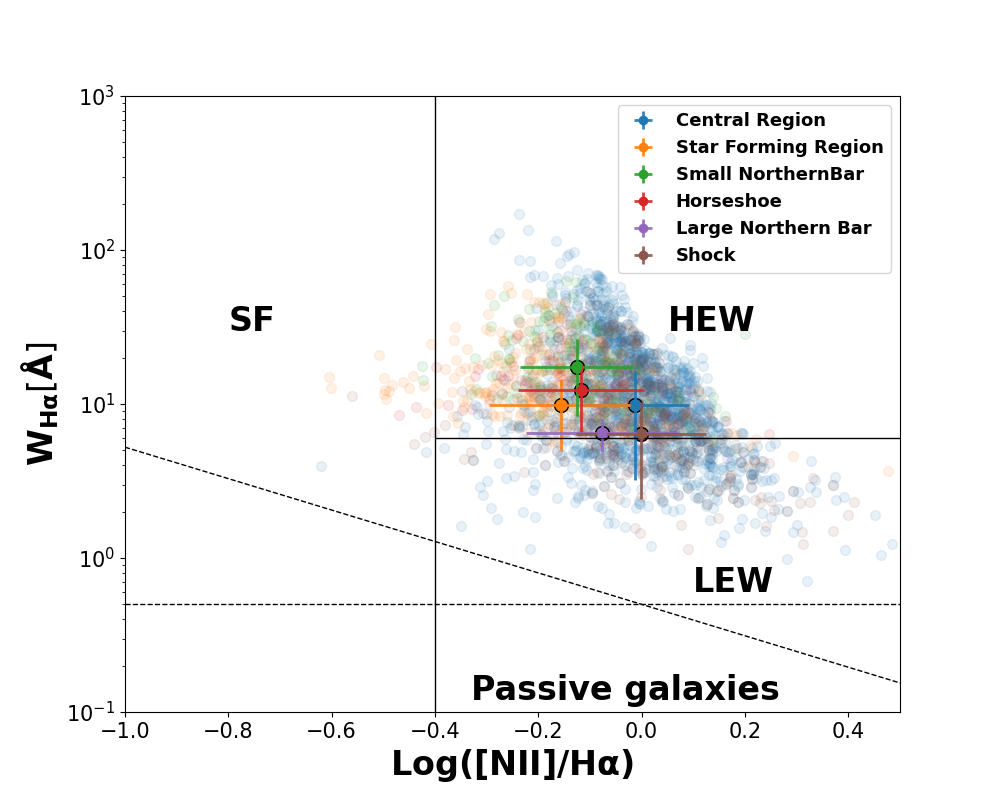}
    \caption{WHAN plot divided by structures. We show the ensemble of points color-coded to each structures studied in detail in $\S$\ref{sec:regionDiscussion}. For each structure, the mean value with a 1-$\sigma$ error is superimposed on the data points.}
    \label{fig:WHANbyRegion}
\end{figure}

Figure \ref{fig:WHANbyRegion} shows the WHAN diagram of the nebula by structures of interest; we have the following structures: the central region, the star-forming region, the small northern filament, the large northern filament, the horseshoe region, and the shock region (refer to figure \ref{fig:combinedImage} for location of structures). In this figure, we have included the mean value and 1-$\sigma$ deviation for each structures. This reveals the regions of interest all lie well within the HEW categorization implying a highly energetic and strong ionizing source for the majority of the spaxels in the structures. However, we note there are some departures from the mean most notably in the shock region which contains more LEW spaxels and in the star-forming region where a handful of pixels fall within the star-forming sequence. The WHAN plots and morphology of these regions are studied individually in the subsequent subsections.

%We begin our detailed analysis with the ICM shock region.

\subsubsection{ICM Shock Structure}
\begin{figure*}
    \gridline{
        \fig{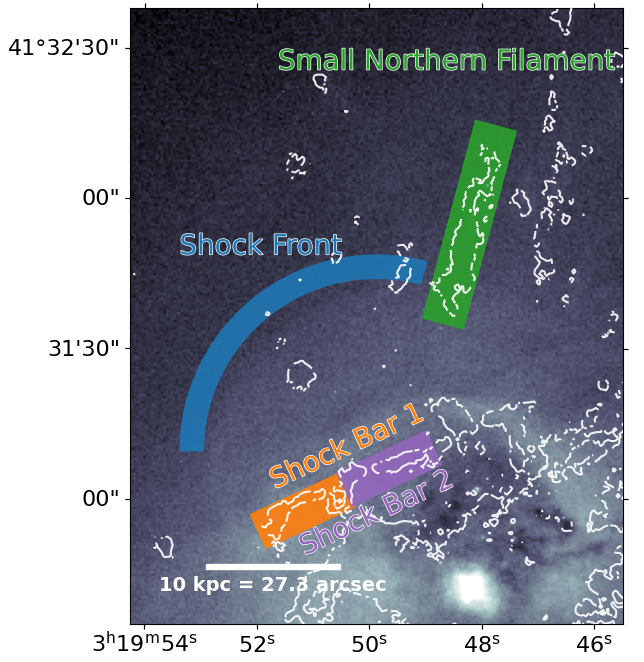}{0.4\textwidth}{(a) Diagram of the shock region in NGC 1275.}
        \fig{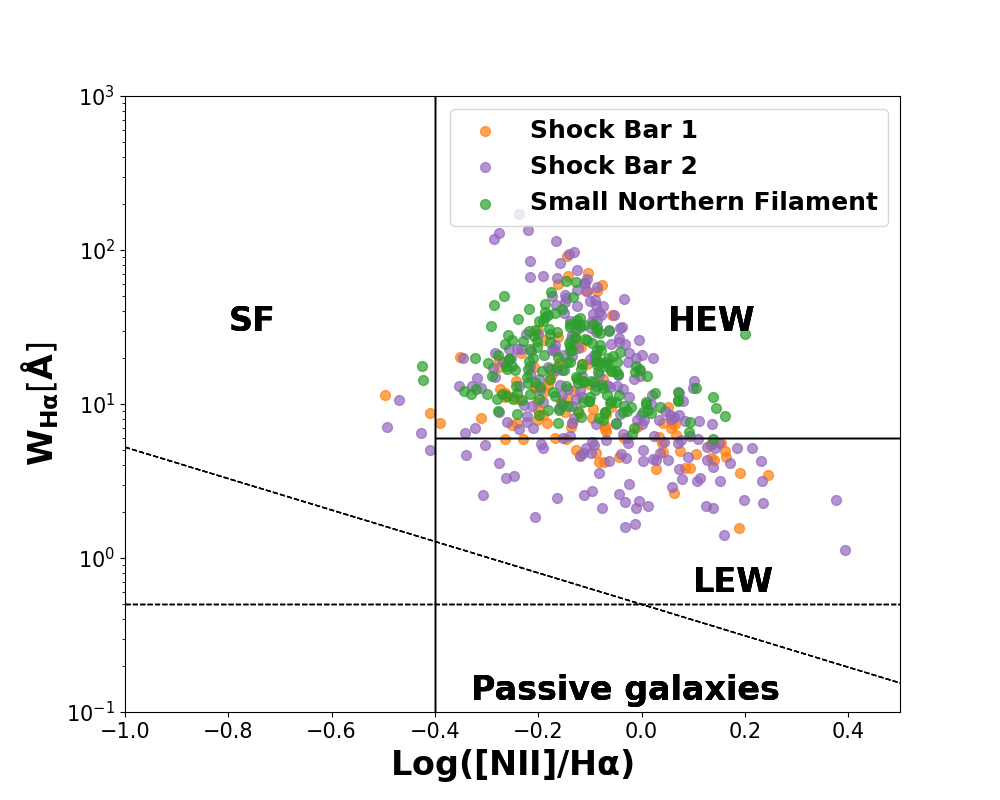}{0.55\textwidth}{(b) WHAN plot for the bar in the shock region.}
    }
    \caption{The left panel is a diagram of the shock region in NGC 1275. The background image is the \textit{Chandra} soft X-ray emission between 0.5-2.0 keV described in $\S$\ref{sec:methodChandra}; the contours highlight the H$\alpha$ emission. We highlight the location of the shock front (blue) as detected in \citealt{fabian_heating_2006}. In the right-hand panel, we show the WHAN plot for the bar in the shock region (orange) and the small northern filament (green) that has not yet gone through the shock.}
    \label{fig:shock}
\end{figure*}

In figure \ref{fig:shock}, we show a diagram of the shock structure using a soft X-ray image between 0.5-2.0 keV to highlight the extent of the ICM along with H$\alpha$ contours added in white. We also added the location of the X-ray shock front as determined by \cite{fabian_heating_2006} in blue, the small northern filament in green, and the structure of the nebula deemed the \textit{shocked bar} in orange and purple.
We divide the \textit{shocked bar} into two distinct regions since the literature suggests there are two distinct filaments in this complex (\citealt{hatch_origin_2006}).
Our hypothesis is that the shocked bar has likely already been affected by the shock front while the small northern filament, with the exception of its most southern part, has not. Therefore, we expect to see the spaxels of each region occupying different spaces in the WHAN diagram. As evidenced on the right-hand side of figure \ref{fig:shock}, these two regions both lie primarily within the HEW section of the diagram indicating that the ionization mechanism of these two regions are not distinct. There is, however, a slight difference since the shock bar contains spaxels well within the LEW section. In fact, a 2-dimensional Kolmogrov-Smirnoff test validates this hypothesis (\citealt{peacock_two-dimensional_1983}; \citealt{fasano_multidimensional_1987}). This indicates that a portion of the shock bar does not require a strong highly energetic ionizing source but rather a weaker highly energetic ionizing source. Additionally, we tested if the ionizing source of the small northern filament changes as a function of distance from the base of the filament, yet we found no change. However, the large scatter in the shocked bar could indicate that the multiphase gas was affected by the passage of the shock. Moreover, figure \ref{fig:colorcoded-seyfert} indicates that the log(\nii{}/H$\alpha$) value remains low and constant over the entire shocked bar.

%Left panel : Chandra image + filaments - show the location shock. 
%Then right panel - BPT diagram
%BPT diagram shows that a strong hard-ioinizing agent is required -- fits nicely with the idea that the shock is responsible in this region for the ionization. Also follows the geometry nicely. 
%\carter{Plot BPT plot color-coded by distance to shock front. Does this show a trend (i.e. do the spaxels closest to the shock front have a higher Halpha/NII fraction?) -- I tried this and no they do not! They have a slightly higher equivalent width on average compared to the points further away but that is it.}

%- figure \ref{fig:shock} shows that the region that has already passed through the shock lies in a different region of the parameter space

%- For small northern filament there is no radial trend 

\subsubsection{Central Region}
\begin{figure*}
    \gridline{
        \fig{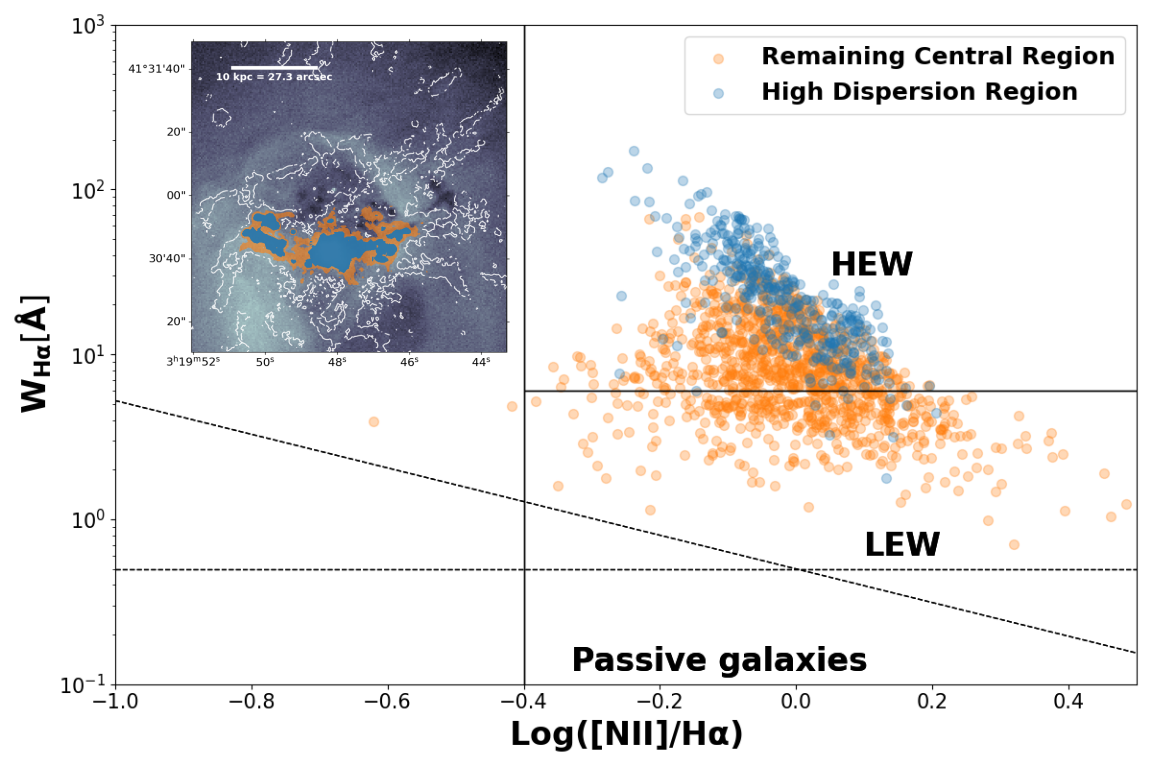}{0.42\textwidth}{(a) Diagram of the central region of the filamentary nebula.}
        \fig{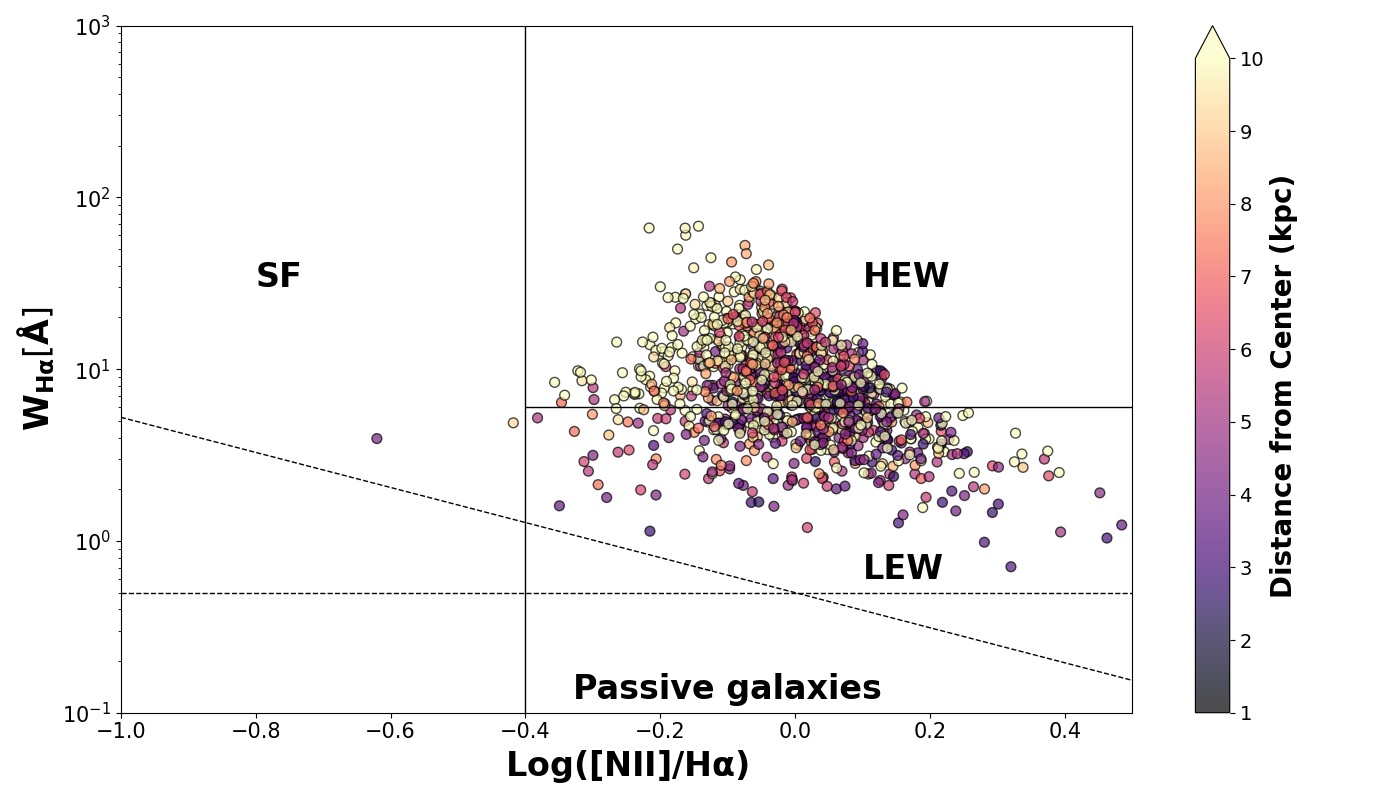}{0.5\textwidth}{(b) WHAN plot color-coded by distance to the AGN.}
    }
    \caption{The left panel is a diagram of the central region of the filamentary nebula in NGC 1275. The background image is the \textit{Chandra} soft X-ray emission, while the white contours highlight the H$\alpha$ emission. We partition the central region into the high dispersion region as initially discovered in \citealt{vigneron_high-spectral-resolution_2024} and the remaining pixels. We show the WHAN diagram for the high dispersion component (blue) and the remaining spaxels in the central region (orange). The right-hand panel shows the WHAN plot color-coded by distance to the AGN.}
    \label{fig:centralRegionBPT}
\end{figure*}

%Flux cut central disk 
Figure \ref{fig:centralRegionBPT} shows the WHAN plot for spaxels in the central region of NGC 1275. Furthermore, we divide the central regions into the high velocity dispersion region\footnote{We adopt the nomenclature of high dispersion region in order to be consistent with \cite{vigneron_high-spectral-resolution_2024}.} initially identified in \cite{vigneron_high-spectral-resolution_2024} and the remaining spaxels in the center. \cite{vigneron_high-spectral-resolution_2024} found that this region is kinematically distinct from the rest of the nebula and is co-spatial with the CO(2-1) disk of molecular gas reported in \cite{nagai_alma_2019}.
Spaxels are considered to be part of the high dispersion region if their flux is above $5\times 10^{-16}$ ergs/s/cm$^{2}$/\AA; this is similar to the d istinction made in \cite{vigneron_high-spectral-resolution_2024}.
Although both groups of spaxels occupy the WHAN plot's Seyfert area, the high dispersion pixels tend to have a higher equivalent width value, indicating a stronger ionizing source. %We note, however, that the distinction between these two populations in the WHAN plot is small; the primary difference remains the stark contrast in velocity dispersion. 

Furthermore, figure \ref{fig:colorcoded-seyfert} shows a structure in the  log(\nii{}/H$\alpha$) values and W$_{H\alpha}$ that corresponds to the high dispersion region; in this region, the values are below unity while in the other parts of the central region are at or above unity.
\cite{vigneron_high-spectral-resolution_2024} proposed that the high dispersion region originates from a distinct mechanism to the rest of the filaments, leading to more turbulent multiphase gas. Here, we see that this high dispersion region is consistent with a region requiring a highly energetic ionization source, perhaps from the AGN's radio jets. Moreover, a 2-dimensional Kolmogrov-Smirnoff test reveals that these two populations are statistically distinct.
In the last panel of the figure, we display each spaxel's position in the WHAN diagram color-coded by their distance from the AGN. There is a slight trend for softer sources the further away the spaxel is from the AGN, but this trend is tenuous. 

Moreover, dedicated studies of the excitation mechanisms of cold molecular gas found in the inner regions of the filamentary structure surrounding NGC 1275 have been carried out. Indeed, multiple near-infrared spectroscopic observations at high-resolution seem to indicate that the molecular gas found in the central 50 pc of the BCG appears to be excited either by shocks from the AGN activity or by X-ray heating from the central source (see \citealt{wilman_nature_2005, scharwachter_kinematics_2013}). Additionally, extended NIFS observations of the inner $900 \times 900\text{ pc}^2$ reveal that powerful AGN-driven outflows could be at the source of the shocks necessary to ionize the inner molecular gas (\citealt{riffel_ionized_2020}). Interestingly, these results could reinforce the idea that the inner regions of the optical filaments close to the central BCG would require a highly ionizing source, most likely through AGN activity, as it can be observed for the central molecular gas.

\subsubsection{Horseshoe Structure}
%Show radio (ML 1.4 GHz - 370 MHz high spatial res where see Northen bubble). 
%-\cite{fabian_relationship_2003} shows horseshoe being dragged up behind bubble. 
%-BPT shows this is well within the Seyfert region implying a stronger hard-ionizing source (figure \ref{fig:horseshoeBPT}).

\begin{figure*}
    \gridline{
        \fig{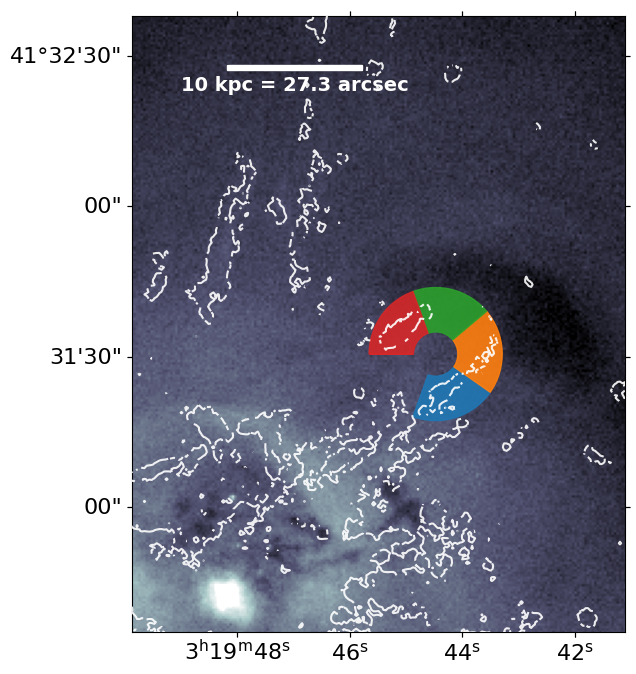}{0.4\textwidth}{(a) Diagram of the horseshoe region in NGC 1275.}
        \fig{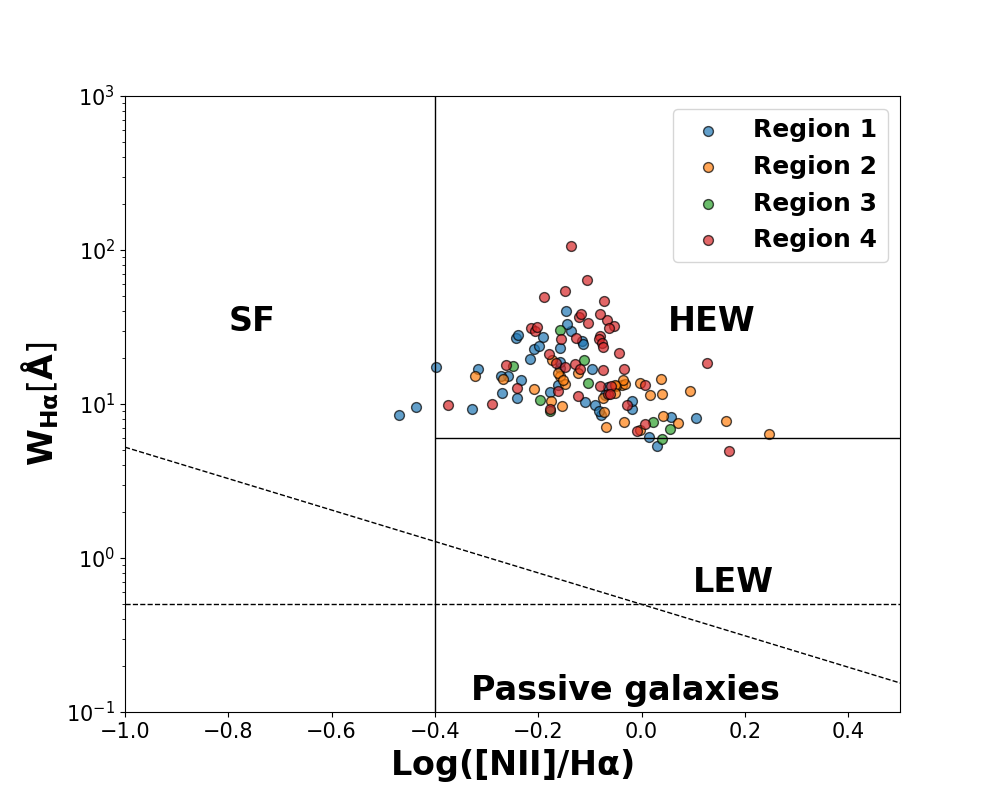}{0.55\textwidth}{(b) WHAN plot for the segmented horseshoe region.}
    }
    \caption{Diagram and WHAN plot of the horseshoe region of the NGC 1275 filamentary nebula. The background image is the \textit{Chandra} soft X-ray emission, while the white contours highlight the H$\alpha$ emission. The horseshoe is segmented into four color-coded sections, capturing the different morphological regions of the area. The colors on the diagram correspond to those in the WHAN plot.}
    \label{fig:horseshoeBPT}
\end{figure*}

Figure \ref{fig:horseshoeBPT} highlights the emission around the horseshoe structure. The colored arcs highlight the horseshoe structure. These regions were chosen to analyze azimuthal trends in the ionization mechanism in different regions of the horseshoe. The results in the WHAN plot (right-hand panel) demonstrate that nearly all spaxels fall within the HEW region of the diagram, implying a source of ionization where the (energy/ionizing event) is higher than for ionizing photons from stars. The spaxels' locations in the diagram are evenly distributed among the different regions. \cite{fabian_relationship_2003} posited that an X-ray bubble created the filamentary structure seen in the figure by dragging out previously ionized gas from the more central regions. While inconclusive, these results agree with that scenario since there is no spatial dependence within the horseshoe for the ionization mechanism.

\subsubsection{Northern Filament}
%- No strong correlation with the distance from the base of the filament (figure \ref{fig:northerFilament}).
%- Implies similar ionization mechanism for all of the filament -- does not appear to be coming from the AGN.
%- Question for Julie/ML: what could be causing this then?
Figure \ref{fig:northerFilament} shows a map of the northern portion of NGC 1275 as seen through the lens of \textit{Chandra} with white contours taken from the H$\alpha$ flux map; we highlight in yellow the location of the large northern filament. On the right-hand panel, we plot each spaxel in the large northern filament on the WHAN plot and color-code them by the distance to the base of the filament. The majority of the points lie within the Seyfert region of the diagram. There are, however, a handful of spaxels lying in the star-forming region which is consistent with results from \cite{canning_star_2010}. 
Importantly, there is no trend in the ionizing mechanism as a function of distance from the base. This lack of a radial trend implies a distributed ionization mechanism is at play in the filament.
Moreover, if the ionization came from the AGN, we would expect evidence for a highly energetic and stronger ionizing source nearer the base, which our data do not show. Therefore, we conclude that AGN photoionization is not responsible for powering the filaments. A uniformly distributed physical phenomenon that is a hard-ionizing source appears to be responsible for the filaments. 

Additionally, figures \ref{fig:colorcoded-seyfert} and \ref{fig:WHANbyRegion} reveal a very small scatter in the equivalent width values in the northern filament compared to all other regions studied. This hints that the ionization mechanism is steady across the entire northern filament.
The homogeneity in values across the northern filament, coupled with the fact that they are coincident with soft X-ray emission, supports the theory posited that the hot X-ray gas is responsible for the ionization of the cool filamentary gas (e.g., \citealt{fabian_energy_2011}; \citealt{werner_feedback_2010}). Previous works have considered that the hot X-ray emitting particles ionize the cool gas; the particles penetrate the cool gas causing a cascade of secondary electrons that act as an ionizing agent for the cool gas (\citealt{ferland_collisional_2009}; \citealt{werner_origin_2014}). The hot ICM particles are potentially able to penetrate the cool gas initially through magnetic reconnection (e.g., \citealt{fabian_energy_2011}). Simulations support the hypothesis demonstrating that collisional excitation and mixing are sufficient to explain the observation line ratios (e.g., \citealt{canning_collisional_2016}).
%Chandra + filaments 

\begin{figure*}
    \gridline{
        \fig{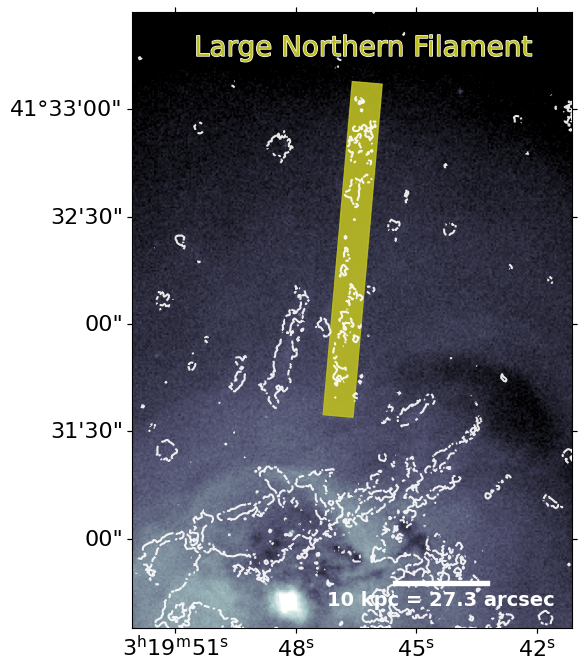}{0.35\textwidth}{(a) Diagram of the larger northern filament.}
        \fig{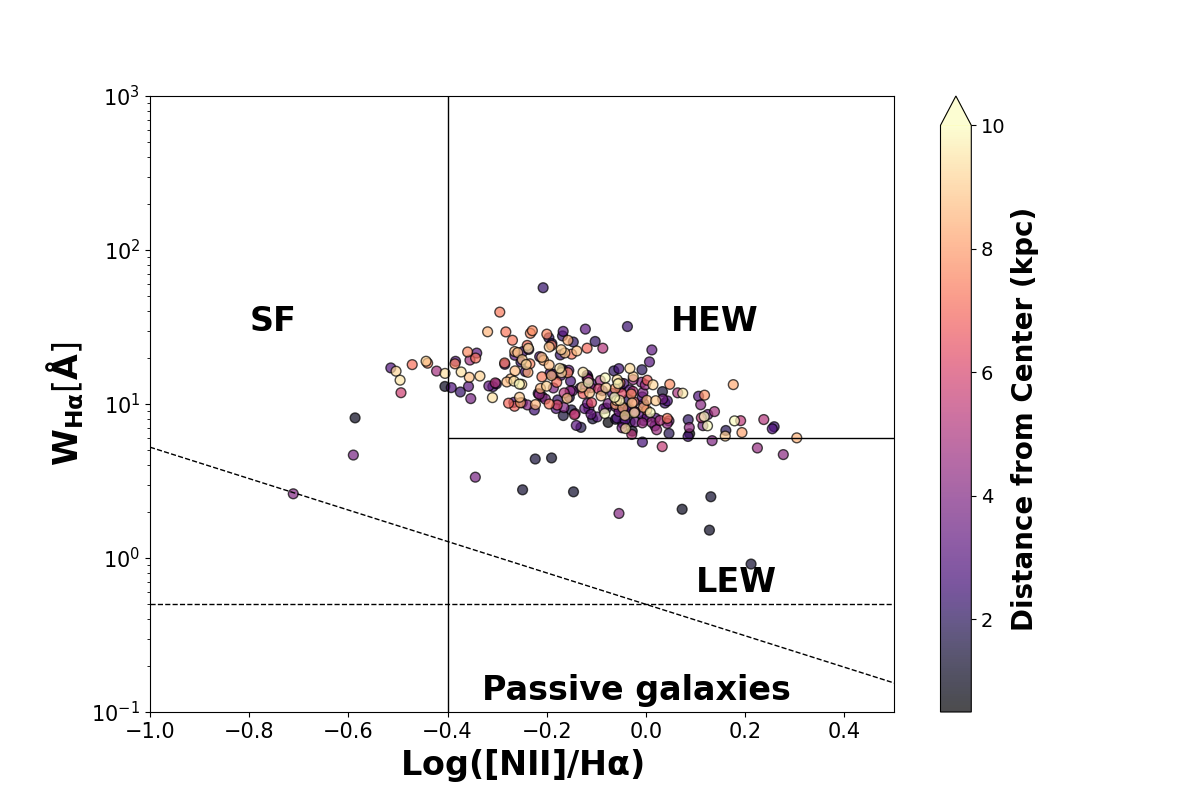}{0.59\textwidth}{(b) WHAN plot color-coded by distance along the filament.}
    }
    \caption{The left-hand plot is a diagram of the larger northern filament. The background image shows the soft X-ray emission captured by \textit{Chandra}, with white contours following the H$\alpha$ emission. The WHAN plot on the right is color-coded by the distance to the bottom of the filament.}
    \label{fig:northernFilament}
\end{figure*}

\subsubsection{Star-Formation in the Blue-Loop}
\begin{figure*}
    \gridline{
        \fig{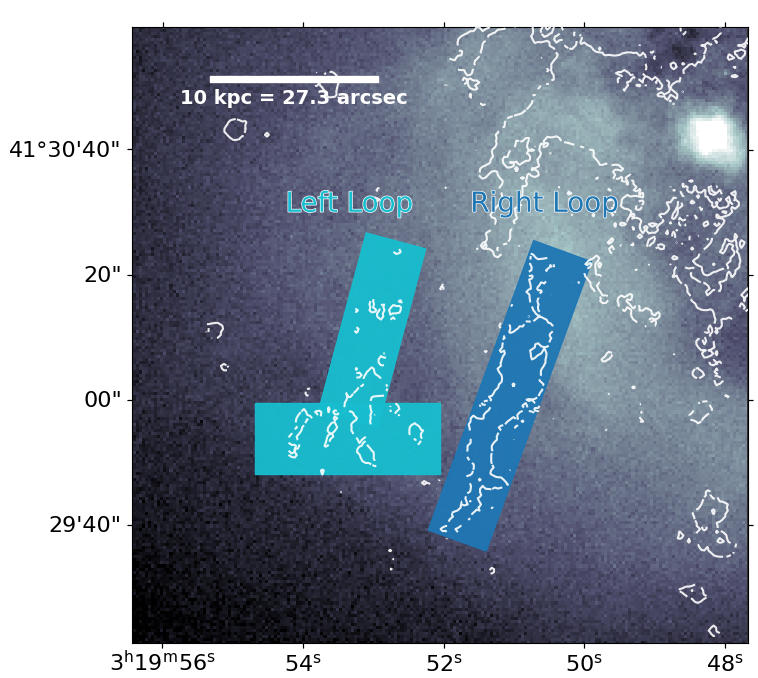}{0.44\textwidth}{(a) Diagram of the Blue Loop region.}
        \fig{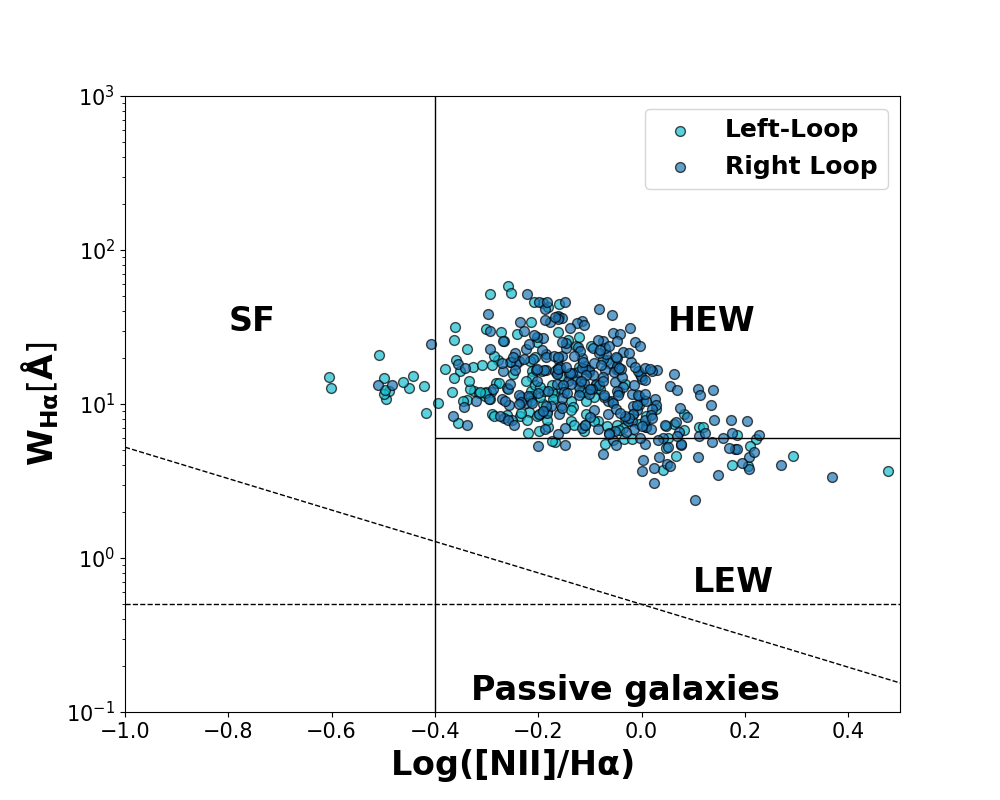}{0.55\textwidth}{(b) WHAN plot color-coded by the left and right loop segments.}
    }
    \caption{In the left panel, we have a diagram of the Blue Loop region. The background image shows the soft X-ray emission captured by \textit{Chandra}, with white contours following the H$\alpha$ emission. The loop is segmented into a left and right loop. The WHAN plot is color-coded by this distinction.}
    \label{fig:blueLoop}
\end{figure*}

%figure that shows star formation 
%Email becky ask the Hubble images
Figure \ref{fig:blueLoop} shows the blue loop as defined by \cite{canning_star_2010}. We further subdivide the blue loop into a left loop and a right loop. The WHAN plot (right-hand panel) shows that the majority of the spaxels lie within the Seyfert region of the diagram, which indicates that a source highly energetic than photoionization is responsible for the ionized gas. However, several spaxels lie within the star-forming region of the diagram, indicating that photoionization is sufficient to explain the ionized gas in these spaxels. This, along with the results from \cite{canning_star_2010}, which show the presence of hot and young stars in the left loop, supports the presence of star-forming regions. 

With regards to star formation throughout the nebula as a whole, there is little indication in the previous literature or from the findings presented here to indicate star formation occuring in traditionally observed regions such as \hii{} regions in the disk of the galaxy. Instead, the regions where the data does hint at star formation are clumpy and in the outer filaments.
Although previous studies on these type of optically-emitting filamentary nebula surrounding BCGs have indicated that a mixture of shocks and star formation can account for the high log(\nii{}/H$\alpha$) values calculated here (e.g., \citealt{mcdonald_optical_2011}), results from the literature indicate that star formation is not likely playing a crucial role in NGC 1275.
%- Coincides with the blue loop from Canning -- mplies star formation (or needs soft ionizing source) -- consistent with our findings in BPT (figure \ref{fig:blueLoop}).

%- Does not require star formation, but it does show that the source could be softer (Cid-Fernandes). Since we see hot/young stars, this is likely the cause (Canning)

%- \oii{} emission in the nebula but not \oiii{} emission.
%https://www.ucolick.org/~simard/phd/root/node21.html
%- Indicates that star formation is not sufficient to ionize regions -- consistent with all previous results.

%\subsection{Overall Trends}
%Recently, several studies have demonstrated the tentative discovery of hidden cooling flows being formed within clusters of galaxies (see \citealt{fabian2022hidden}, \citealt{fabian2023ahidden}, \citealt{fabian2023bhidden}). These analyses suggest that the energy injection generated by the AGN within the ICM to prevent its cooling would in fact not be as effective as previously determined. The ICM would thus be capable of cooling leading to the presence of significant molecular gas structures within the cluster environment. They argue that these cooling flows would be `hidden` within X-ray wavebands due to obsfucation by dust and photoelectrically-absorbing cold clouds resulting in absorbed emission being transfered to the infrared domain (e.g. \citealt{white1991discovery}, \citealt{johnstone1992spectral}, \citealt{fabian1994asca}). Therefore, intense cooling of the ICM around BCGs of galaxy clusters could prevent the detection of these cooling flows as a result of low temperature and emissivity (\citealt{fabian2023ahidden}). Measurements on a selected sample of galaxy clusters indicate that cooling flows of $\sim 30 - 100$ $M_{\odot}\text{.yr}^{-1}$ could potentially be found within them and incidently within the Perseus cluster of galaxies (\citealt{fabian2022hidden}). 

\section{Summary}
The central galaxy of the Perseus cluster, NGC 1275, plays host to an emission-line nebula visible clearly in optical wavelengths. The nebula is filamentary, ranging from the inner kiloparsecs to several dozens of kiloparsecs outwards.
While previous studies indicated a lack of star formation in most of the nebula (e.g., \citealt{canning_star_2010}), no robust and systematic study on the ionization mechanism of different regions across the entire nebula existed. In this paper, we use three SITELLE observations covering the majority of the optical regime to study the ionization mechanisms at play in the nebula. Though most regions of the nebula do not contain any \oiii{}, a finding consistent with \cite{hatch_origin_2006}, we demonstrate that \oiii{}$\lambda$5007 is present in the core of the emission-line nebula. Moreover, we find a slight radial trend in the strength of the ionizing source as a function of distance from the center of the cluster.

Due to the lack of detected \oiii{} emission in the extended filaments, we rely on the WHAN diagnostic plots (e.g., \citealt{cid_fernandes_alternative_2010}; \citealt{cid_fernandes_comprehensive_2011}), which do not require \oiii{} to distinguish between the ionization mechanisms.
We find the following:
\begin{itemize}
    \item the large scatter in the shock bar implies that the shock has affected the ionization of multiphase gas.
    \item the high dispersion region in the core requires a highly energetic ionizing source. This source may be from high-energy photoionization, shocks, or the cooling ICM. 
    \item the ionization in the horseshoe structure is homogeneous throughout.
    \item the ionization mechanism in the Northern filament is a strong and highly energetic-ionizing source that is uniform across the filament.
    \item there is little star formation occurring throughout the nebula.
    \item the AGN does not play a critical role in the ionization of the filaments. While the exact extent of the AGN's role in the central regions of the nebula is not fully constrained, it is unlikely that it plays a crucial role beyond its immediate surroundings.
\end{itemize}

In this work, we have demonstrated the complexity of the ionization process throughout the different structures in the NGC 1275 filamentary nebula.
Additionally, we have utilized this dataset to map the optical emission of the entire nebula and bolstered the argument that some hard ioinzing source other than the central AGN is responsible for the ionization of the nebula.
While illuminating, there remains much work to disentangle the different potential ionization processes.

\software{
\luci{} \citealt{rhea_crhea93luci_2021}; 
\texttt{ds9} \citealt{joye_new_2003}; \texttt{ndtest} \citealt{li_ndtest_2023}; \texttt{matplotlib} \citealt{hunter_matplotlib_2007}; \texttt{scipy} \citealt{virtanen_scipy_2020}; \texttt{numpy} \citealt{harris_array_2020}; \texttt{tensorflow} \citealt{abadi_tensorflow_2015}; \texttt{astropy} \citealt{astropy_collaboration_astropy_2018} and \citealt{astropy_collaboration_astropy_2022}; \texttt{astroalign} \citealt{beroiz_astroalign_2020}
}

\section*{Acknowledgements}
The authors would like to thank the Canada-France-Hawaii Telescope (CFHT) which is operated by the National Research Council (NRC) of Canada, the Institut National des Sciences de l'Univers of the Centre National de la Recherche Scientifique (CNRS) of France, and the University of Hawaii. The observations at the CFHT were performed with care and respect from the summit of Maunakea which is a significant cultural and historic site.
C.L. R. acknowledges financial support from the physics department of the Universit\'e de Montr\'eal, the MITACS scholarship program, and the IVADO doctoral excellence scholarship.
J. H.-L. acknowledges support from NSERC via the Discovery grant program, as well as the Canada Research Chair program.
MLGM acknowledges financial support from the grant CEX2021-001131-S funded by MCIU/AEI/ 10.13039/501100011033, from the coordination of the participation in SKA-SPAIN, funded by the Ministry of Science, Innovation and Universities (MCIU), as well as NSERC via the Discovery grant program and the Canada Research Chair program.
M.M. acknowledges support from the Spanish Ministry of Science and Innovation through the project PID2021-124243NB-C22. This work was partially supported by the program Unidad de Excelencia Mar\'ia de Maeztu CEX2020-001058-M.
N. W. is supported by the GACR grant 21-13491X.
ACE acknowledges support from STFC (ST/T000244/1, ST/X001075/1).

\bibliography{Perseus}
\bibliographystyle{aasjournal}

\appendix 

\section{Kinematics}\label{app:kinematics}
In this section, we show the kinematic fits (i.e., velocity and velocity dispersion) as calculated from SN3, SN2, and SN1.
The SN2 maps follow the SN3 maps closely. However, the SN1 kinematic maps suffer from the fact that the OII-doublet is not resolved. These results do not effect the flux in a meaningful way.
%Although not shown since the maps suffer from reduced signal-to-noise, the kinematic maps from SN1 and SN2 are identical in regions of sufficient SNR to resolve the strong emission lines in those cubes. 
\begin{figure*}
    \centering
    \gridline{
            \fig{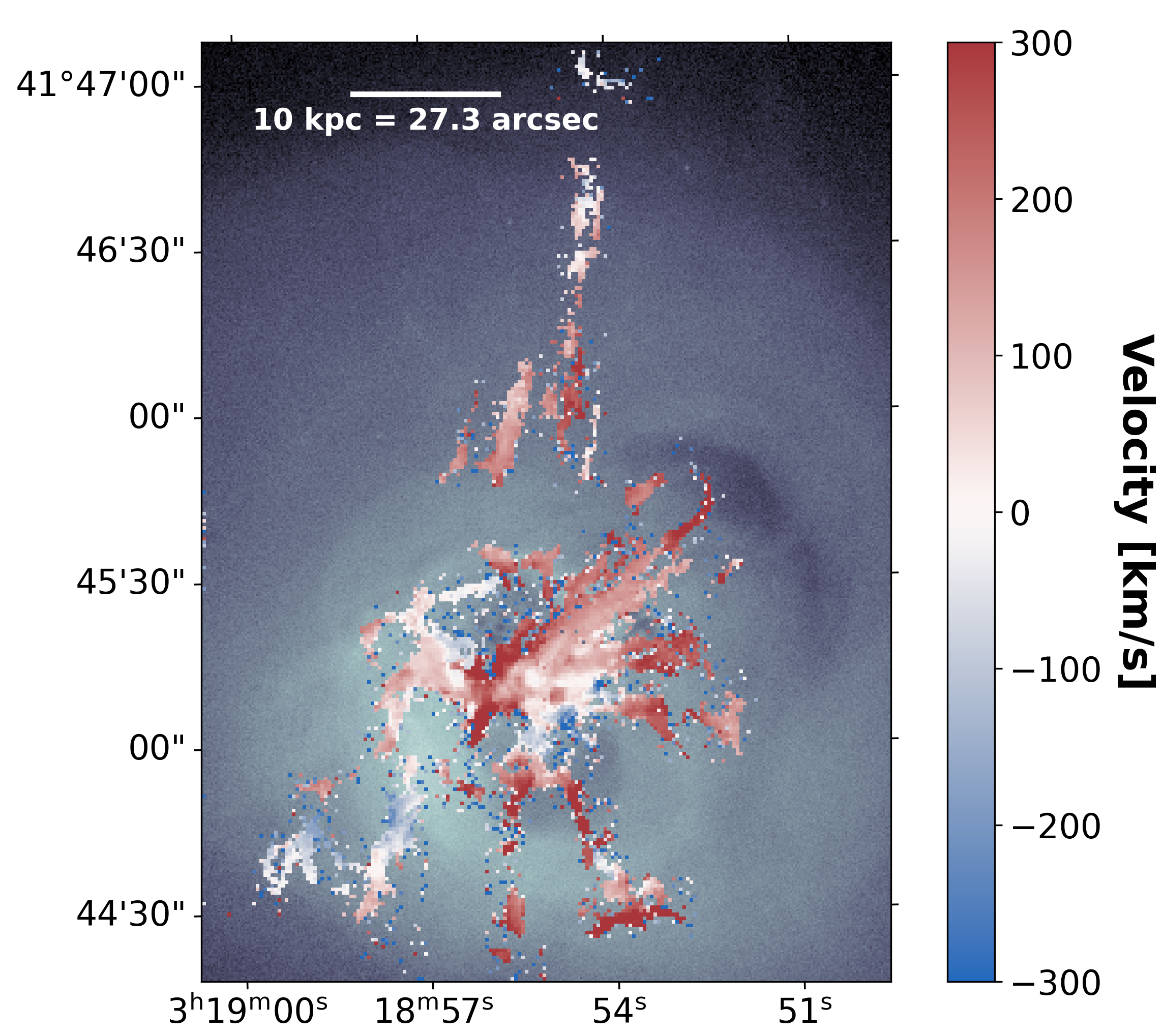}{0.45\textwidth}{(a) Velocity}
            \fig{Halpha_Broadening_masked.png}{0.45\textwidth}{(b) Velocity dispersion}
            }
    \caption{Left: SN3 velocity map; Right: SN3 broadening map. Although these findings were initially reported in \citet{gendron-marsolais_revealing_2018}, we include the \luci{}-based fits here for posterity. We note that globally the values are similar to those calculated previously.}
    \label{fig:kinematics}
\end{figure*}

\begin{figure*}
    \centering
    \gridline{
            \fig{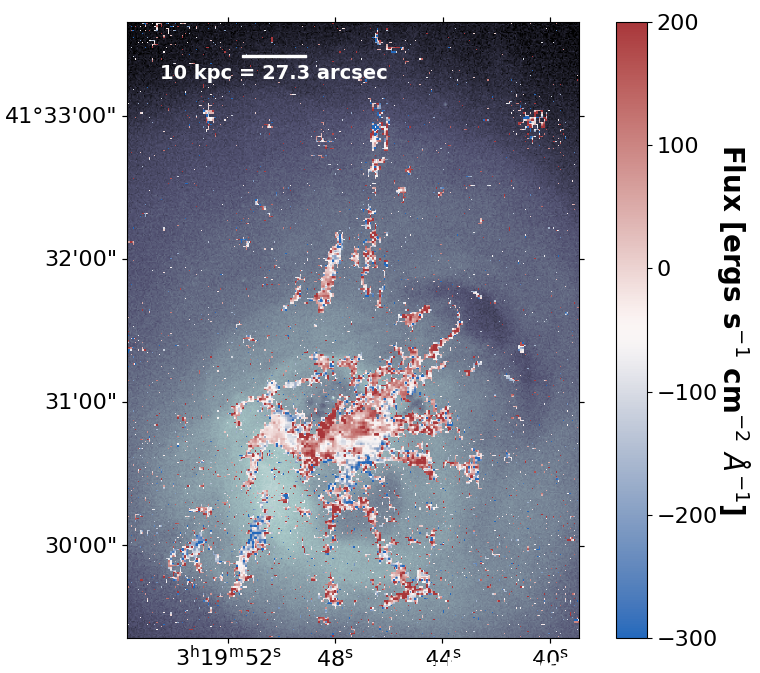}{0.45\textwidth}{(a) Velocity}
            \fig{Hbeta_Broadening_masked.png}{0.45\textwidth}{(b) Velocity dispersion}
            }
    \caption{Left: SN2 velocity map; Right: SN2 broadening map. The values reported for the SN2 data follows the SN3 data closely.}
    \label{fig:kinematics}
\end{figure*}

\begin{figure*}
    \centering
    \gridline{
            \fig{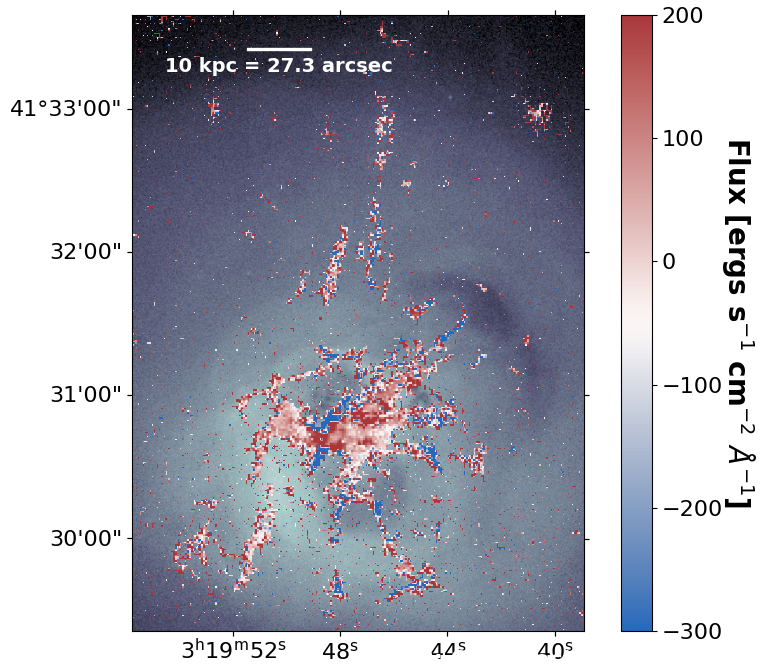}{0.45\textwidth}{(a) Velocity}
            \fig{OII_Broadening_masked.png}{0.45\textwidth}{(b) Velocity dispersion}
            }
    \caption{Left: SN1 velocity map; Right: SN1 broadening map. While the velocity values of the blended \oii{} doublet follow the values reported for SN2 and SN3 closely, the broadening values differ. Instead, they are uniformly between 100 and 120 km/s. This is likely due to the fact that the two lines are completely blended.}
    \label{fig:kinematics}
\end{figure*}

\section{Fit Commands}\label{app:fitCommands}
In this section we show the three \luci{} commands used to fit each filter.
\subsection{Command for fitting SN3}
\begin{python}
vel_map, broad_map, flux_map, chi2_fits = cube.fit_cube(
        ['Halpha', 'NII6548', 'NII6583','SII6716', 'SII6731'],  # lines
        'sincgauss',   # fit function
        [1,1,1,1,1],  # velocity relationship 
        [1,1,1,1,1],  # sigma relationship
        1000, 1500 ,200, 1000, # Bounds on pixels coordinates
        bkg=bkg_sky, # Apply background subtraction
        binning=3,  # Binning 
        n_threads=30,  # Number of threads to run
        uncertainty=True,
        spec_min=15000, spec_max=15400
    )
\end{python}

\subsection{Command for fitting SN2}
\begin{python}
vel_map, broad_map, flux_map, chi2_fits = cube.fit_cube(
        ['Hbeta', 'OIII4959', 'OIII5007'],
        'sincgauss',
        [1,1,1],
        [1,1,1],
        800, 1300 ,700, 1500,
        binning=3,
        bkg=bkg_sky, 
        n_threads=12,
        uncertainty = True,
        spec_min=19750, spec_max=21500,
    )
\end{python}

\subsection{Command for fitting SN1}
\begin{python}
vel_map, broad_map, flux_map, chi2_fits = cube.fit_cube(
        ['OII3726', 'OII3729'], 
        'sincgauss',
        [1,1], [1,1],
        800, 1300 ,700, 1500,
        binning=3,
        bkg=bkg_sky, 
        n_threads=12,
        uncertainty = True
    )
\end{python}



\end{document}